\documentclass[3p,times]{elsarticle}

\usepackage{ecrc}

\volume{00}

\firstpage{1}

\journalname{Computer Networks}

\runauth{}

\jid{procs}

\jnltitlelogo{Computer Networks}

\CopyrightLine{2023}{Published by Elsevier Ltd.}

\usepackage{amssymb}
\usepackage[figuresright]{rotating}

\usepackage{url}
\usepackage{amsmath,amssymb,amsfonts}
\usepackage{algorithmic}
\usepackage{graphicx}
\usepackage{textcomp}
\usepackage{xcolor,colortbl}
\usepackage{mathtools}
\usepackage{todonotes}
\usepackage{booktabs}
\usepackage{float}
\usepackage{hyperref}
\usepackage{enumitem}
\usepackage[export]{adjustbox}
\usepackage{pifont}
\def\BibTeX{{\rm B\kern-.05em{\sc i\kern-.025em b}\kern-.08em
    T\kern-.1667em\lower.7ex\hbox{E}\kern-.125emX}}
\definecolor{Gray}{gray}{0.8}
\definecolor{LigthGray}{gray}{0.95}

\definecolor{CM10}{HTML}{008800}
\definecolor{CM09}{HTML}{00AA00}
\definecolor{CM08}{HTML}{00CC00}
\definecolor{CM07}{HTML}{00FF00}
\definecolor{CM02}{HTML}{A0FFA0}
\definecolor{CM01}{HTML}{C0FFC0}
\definecolor{CM00}{HTML}{EEFFEE}

\newcommand{\figref}[1]{Figure~\ref{#1}}
\newcommand{\secref}[1]{Section~\ref{#1}}
\newcommand{\tabref}[1]{Table~\ref{#1}}

\newcommand\mynobreakpar{\par\nobreak\@afterheading}

\usepackage[linesnumbered,ruled,vlined]{algorithm2e}
\usepackage{multirow}
\usepackage{array}
\usepackage{float}
\usepackage{acronym}
\usepackage{amsmath}
\usepackage{caption}
\usepackage{subcaption}

\newcolumntype{F}[1]{>{\columncolor{white}\raggedright}p{#1}}
\newcolumntype{S}[1]{>{\raggedright\arraybackslash\columncolor{LigthGray}}p{#1}}
\newcolumntype{L}{>{\columncolor{Gray}}l}
\newcolumntype{Q}{>{\columncolor{white}}l}
\newcolumntype{a}{>{\columncolor{Gray}}c}
\newcolumntype{P}[1]{>{\centering\arraybackslash\columncolor{white}}p{#1}}
\newcolumntype{Z}[1]{>{\centering\arraybackslash\columncolor{LigthGray}}p{#1}}
\newcolumntype{X}{>{\columncolor{white}}c}
\newcolumntype{Y}{>{\columncolor{white}}l}
\newcolumntype{R}{>{\columncolor{white}}r}
\newcolumntype{O}[1]{>{\raggedleft\arraybackslash\columncolor{white}}p{#1}}
\newcolumntype{M}[1]{>{\centering\arraybackslash\columncolor{white}}p{#1}}
\newcolumntype{N}[1]{>{\raggedright\arraybackslash\columncolor{white}}p{#1}}

\makeatletter
 \let\old@ps@headings\ps@headings
 \let\old@ps@IEEEtitlepagestyle\ps@IEEEtitlepagestyle
 \def\confheader#1{%
 \def\ps@headings{%
 \old@ps@headings%
 \def\@oddhead{\strut\hfill#1\hfill\strut}%
 \def\@evenhead{\strut\hfill#1\hfill\strut}%
 }%
 \def\ps@IEEEtitlepagestyle{%
 \old@ps@IEEEtitlepagestyle%
 \def\@oddhead{\strut\hfill#1\hfill\strut}%
 \def\@evenhead{\strut\hfill#1\hfill\strut}%
 }%
 \ps@headings%
 }
\makeatother

\confheader{%
\textit{Submitted to The International Journal of Computer and Telecommunications Networking}}

\begin{document}

\begin{frontmatter}



\dochead{}
%% e.g. \dochead{17th International Conference on Dynamical Processes in Excited States of Solids}

\title{NetTiSA: Extended IP Flow with Time-series Features for Universal Bandwidth-constrained High-speed Network Traffic Classification}


\author[ctu]{Josef Koumar}
\author[cesnet]{Karel Hynek}
\author[ctu]{Jaroslav Pešek}
\author[cesnet]{Tomáš Čejka}

\address[ctu]{Czech Technical University in Prague, Faculty of Information Technology, Thákurova 9, 160 00 Prague 6 }
\address[cesnet]{CESNET a.l.e., Generála Píky 430/26, 160 00 Prague 6 }

\begin{abstract}
    Network traffic monitoring based on IP Flows is a standard monitoring approach that can be deployed to various network infrastructures, even the large IPS-based networks connecting millions of people. Since flow records traditionally contain only limited information (addresses, transport ports, and amount of exchanged data), they are also commonly extended for additional features that enable network traffic analysis with high accuracy. Nevertheless, the flow extensions are often too large or hard to compute, which limits their deployment only to smaller-sized networks. This paper proposes a novel extended IP flow called \textit{NetTiSA (Network Time Series Analysed)}, which is based on the analysis of the time series of packet sizes. By thoroughly testing 25 different network classification tasks, we show the broad applicability and high usability of \textit{NetTiSA}, which often outperforms the best-performing related works. For practical deployment, we also consider the sizes of flows extended for \textit{NetTiSA} and evaluate the performance impacts of its computation in the flow exporter. The novel feature set proved universal and deployable to high-speed ISP networks with 100\,Gbps lines; thus, it enables accurate and widespread network security protection. 
\end{abstract}

\begin{keyword}
 time series \sep unevenly spaced time series \sep time series analysis \sep classification \sep computer networks \sep machine learning \sep IP flow \sep flow exporter
\end{keyword}

\end{frontmatter}

% \linenumbers


\section{Introduction}
\label{introduction}
    Network monitoring plays a crucial role in the overall computer security management. Compared to protections (such as AntiVirus software) deployed on the end devices, the network-based incident detection and prevention systems can protect infrastructure against users' sloppiness, policy violations, or (at worst) intentional attacks from the inside. However, maintaining network security has become increasingly challenging in recent years due to mass traffic encryption and consequent reduced visibility. The encryption of TLS certificates by TLS1.3~\cite{rfc8446}, deployment of encrypted DNS~\cite{rfc8484}, or Encrypted Client Hello proposal~\cite{ietf-tls-esni-16} removed the few-remaining information essential for effective threat detection. Therefore, if possible, the security managers are forced to deploy intermediate proxies to decrypt the traffic~\cite{enisa2019} and inspect it via Deep Packet Inspection (DPI) tools such as Suricata\footnote{\url{https://suricata.io}}. In such cases, the deployment of the intermediate proxy is much more intrusive than sending domain names and certificates in plaintext.  The DPI combined with the proxy is an efficient solution for mid-sized restricted networks. However, it does not scale up to large provider-based networks, where threat detection is also desired~\cite{Aqil2017}. 

    Internet Service Providers (ISP) need to perform network monitoring and cybersecurity threat detection to force internal policies, protect their infrastructure, and prevent overloading of their lines to maintain service~\cite{Aqil2017}. In ISP deployment, the use of the intermediate proxy is unthinkable. It would outrage the consumers due to absolute payload availability; moreover, it is not feasible to process such a large amount of traffic transmitted over multiple 100Gbps backbone lines with DPI. Therefore, large-scale infrastructures are often monitored using the flow-based approach, where each communication is represented in form flows. The flows are collected at the observation points and are transmitted to the flow collector using a flow-export protocol such as the \textit{Internet Protocol Flow Information Export (IPFIX)~\cite{rfc7011}} or NetFlow Version 9~\cite{rfc3954}.

    Even though there are not any rigid specifications of flow, traditionally, it contains only IP addresses, transport ports, and the amount of transferred data~\cite{Hofstede2014}. Raw flows have been previously successfully used in volumetric attack detection such as DDoS~\cite{Sperotto2010}. Detection of network communication that cannot be distinguished volumetrically is with traditional flows challenging and often requires additional data source or active probing confirmation~\cite{doh_kamil} which increases the load on the already resource-constrained monitoring hardware. To avoid active probing or information gathering from external sources, researchers perform various flow extensions to provide additional information that enables accurate classification of network traffic~\cite{9796558,cic_bell_dns_2021_article,luxemburk2022fine}. 

    There are many detection proposals in the field of network security that use extended flows combined with multiple techniques, including machine learning. However, these approaches mainly focus on model accuracy and not practical usability; thus, we can see almost no real-world deployment of these approaches. The feasibility of deployment is usually not a concern. For example, they extend flows for features that are too resource-intensive for computation, allowing their usage only in an offline manner as discussed by Jerabek et al.~\cite{doh_kamil,montazerishatoori2020detection}. 

    Other approaches extend the flows for too much data, such as a sequence of packet lengths and times~\cite{LuxemburkQUIC}, sequence of packets burst and times~\cite{sblt_our_paper}, initial-data-packet content~\cite{plny2023decrypto}, or simply too many individual features that increase the flow size several times as in the case of CICFlowMeter\footnote{\url{https://github.com/ahlashkari/CICFlowMeter}}. The size of the flow telemetry matters for deployment since it occupies bandwidth that could be otherwise commercialized, and it significantly increases the performance requirements on the monitoring infrastructure that needs to process more data in a given time.  On the contrary, the detectors with small feature vectors (such as~\cite{velasco2021efficient}) are not universal; their specifically tailored features were proved to work on a single task, but other tasks were not evaluated. Thus, there is still a need for small but universal features, easily extractable from network traffic that would finally enable encrypted network traffic classification at a scale.

    Motivated by the previous statements, we explored the possibility of creating universal and small-enough feature vectors that could be deployed to high-speed network monitoring lines, thus enabling flow-based encrypted traffic monitoring and analysis at a scale. Therefore, we used \textit{Time Series Analysis (TSA)} of \textit{Single Flow Time Series (SFTS)}~\cite{KoumarUSTS} that can capture long-lasting and short flows and is applicable in multiple network classification tasks~\cite{koumar2023}. 

    Each flow is internally represented as a time series of network packets, i.e., one datapoint of SFTS describes a size of packet payload with a position in a time series defined by transmission time. Using the TSA of these time series, we generated a set of novel features called \textit{NetTiSA (Network Time Series Analyzed flow)}, exported as a novel extended flow. Furthermore, the NetTiSA feature computation is highly optimized and can be computed in a streamwise manner without the necessity of saving the time series into operational memory. The feature set describes time dependencies between packets, packet sequences, distribution of packets, and behavior of packets. 
    
    To validate the usability of the NetTiSA features, we evaluated them using well-known network traffic monitoring tasks, well-known published datasets, and machine learning (ML). According to our validation, the proposed NetTiSA features are usable in both binary and multiclass network traffic classification problems. Moreover, our approach outperforms a majority of best-performing related works of classification by IP flows on the same datasets while requiring minimal bandwidth for the telemetry compared to related works.

    The contributions of our work can be summarized as follows:

    \begin{itemize}
        \item[--] We proposed a novel network traffic analysis approach that uses time series analysis inside the IP flow exporter to generate 13 well-known and novel features exported in the extended IP flow called \textit{NetTiSA flow}. Furthermore, seven more features are computed from the NetTiSA flow before classification, resulting in a feature set with 20 features called \textit{Enhanced NetTiSA flow}.   
        \item[--] We detect multiple potential network threats with \textit{Enhanced NetTiSA flow} as input for ML. Our ML models achieved high Accuracy and F1 scores, exceeding the best results from the related works using the same datasets. Threats include Botnet, Cryptomining, DoH, (D)DoS, Malicious DNS, Intrusion in IDS, IoT Malware, Tor,  and VPN.
        \item[--] We also perform multiclass classification with \textit{Enhanced NetTiSA flow} as input of ML. Our ML models achieved high Accuracy and F1 scores that exceeded the best previous results from relevant works using the same datasets. The multiclass classification includes Botnet, IDS, IoT Malware, Tor, and VPN.  % , and QUIC
        \item[--] Our approach achieves the best results with smaller network telemetry compared to related works, is universally applicable to multiple classification problems, and enables high-speed network classification.
        \item[--] The implementation of the \textit{NetTiSA flow} is publicly available in open source IP flow exporter ipfixprobe\footnote{\url{https://github.com/CESNET/ipfixprobe}}. The implementation is suitable for high-speed ISP networks and can process 100\,Gbps of network traffic. Furthermore, monitoring by the \textit{NetTiSA flow} is currently deployed into ISP network CESNET2\footnote{The Czech Educational and Science Network}. 
        \item[--] We publish datasets containing the \textit{Enhanced NetTiSA flow} features created from 15 well-known publicly available datasets. The created datasets are available on Zenodo \cite{OUR_DATASET}. 
    \end{itemize}

    This paper is divided as follows: \secref{related_works_section} summarizes the related work on the flow-based classification of network traffic. \secref{sec:dataset_description} describes the selected datasets and network traffic classification usecases. \secref{tsa_in_flow_exporter} provides information on time series analysis concepts and motivation and describes a novel approach to time series analysis in the IP flow exporter. \secref{exported_features} provides a complete description of the features exported in the novel extended IP flow. \secref{classification_section} describes the complete classification pipeline and presents the results of using the \textit{NetTiSA} flow for classification. \secref{sec:telemetry_comparison} analyzes the proposed feature vector size and compares it with related works. \secref{sec:implementation} provides information about the impact of NetTiSA features calculation. We present a high-speed C++ implementation of the \textit{NetTiSA} flow inside the IP flows exporter ipfixprobe, and compare performance with related flow extensions. \secref{conclusion_section} concludes this paper.

\section{Related Works} \label{related_works_section}
    The flow-based network monitoring is an essential approach for telemetry acquisition in large network infrastructures and maintaining security. The flow-based intrusion detection systems deploy a variety of classifiers and detectors of malicious communications. Naturally, the design of these detectors strongly depends on the information available in the flow. Since flow record has a relatively loose definition~\cite{Hofstede2014}, records can contain almost any data that can be extracted from the communication. Modern flow export protocols such as IPFIX or NetFlow Version 9 support templating mechanisms so that the users can define their own data structures transferred inside flow records. Nowadays, sending variable length lists of common datatypes (such as uint32) or variable length byte arrays is also possible. It is thus no wonder that the flows are often extended for various information that is helpful in intrusion detection.  We can divide the flow extensions into three directions:  1.\,Extension for extracted unencrypted information, 2.\,Extension for packet sequences, and 3.\,Extension for precomputed features. These approaches are further described in the following sections.

    \subsection{Extension for extracted unencrypted information}
        Exporting unencrypted information from the packet payload is one of the essential functionalities of various flow exporters. It is common to export domain names transferred in DNS packets, information from HTTP headers and TLS handshakes, or simply just part of the payload. These payload fields enable a DPI detection approach even with the flow monitoring infrastructure~\cite{velan2015survey}. However, each detection and classification task requires the extraction of different fields from the packets; thus, multiple flow extensions are required in practice, resulting in large telemetry records. 

        % foremski2014dns,
        
        Due to its reliability, the unencrypted information extension is a popular data source in flow-based monitoring; however, it is unusable with encrypted traffic. In recent years, we have seen a trend in traffic encryption and novel privacy-preserving protocols. For example, plain text DNS is being replaced by its encrypted versions, such as DNS over HTTPS (DoH) and DNS over TLS (DoT)~\cite{rfc8484,rfc7858,garcia2021large}. Privacy-sensitive information in TLS handshakes is already sometimes encrypted with the Encrypted Client Hello extension~\cite{tsiatsikas2022measuring}, which is challenging for traffic classification~\cite{shamsimukhametov2022encrypted}. Since most of the traffic nowadays is encrypted, the importance of unencrypted information extraction is rapidly decreasing, and it is being replaced by other types of flow extensions targeting encrypted traffic analysis.
%%%%    
    \subsection{Extension for raw packet information}
        The extension of flows for packet sequences embeds the raw packet-level information about ongoing connections into the flows. Typically, flows are extended for a \textit{Sequence of Packet Lengths and Times (SPLT)} that aims to export the first $n$ packets of flow as a sequence of packet lengths (or payload lengths), directions, and inter-packet times \cite{ACETO2021102985,luxemburk2022fine,lopez2017network}. Each network classifier can then process the SPLT differently to maximize its accuracy. For example, network classifiers proposed by Vekshin et al.~\cite{vekshin2020doh} or Hynek et al.~\cite{hynek2020refined} process SPLT by additional feature extraction (min, max, median of the sequence), which is then used by ML algorithms. 

        Other approaches feed SPLT directly into the ML algorithm. In their study, Luxemburk et al.~\cite{luxemburk2022fine} used SPLT of length 30 packets for fine-grained network classification with a large number of labels. They use information from SPLT (packet lengths, directions, and times) as three separate sequences and process them with a Convolutional neural network (CNN) with 1D convolutions. The proposed approach achieved great results in classification since the convolutions extracted discriminatory features themselves. Moreover, they also showed that tree-based ML algorithms can perform with similar accuracy even if raw SPLT is used as an input.

        Another use of SPLT was proposed by Chen et al.~\cite{chen2017seq2img} and also by Shapira et al.~\cite{shapira2019flowpic}. Both approaches embed flows with SPLT sequence into a 2D image, which is then used by CNN with 2D convolutions. The biggest difference between both approaches is in the size of the used input images. Chen et al.~\cite{chen2017seq2img} uses SPLT for only the first ten packets and achieves a high accuracy of 88.42\% on the classification of network applications. Shapira et al.~\cite{shapira2019flowpic} deals with VPN detection tasks and uses SPLT of very long size. Their proposed detector achieved a 99.7\%  of accuracy; however, it requires $1400 x 1400$ images as the input, which means that the size of each flow with SPLT sequence is at least 7.84 megabytes. Such large telemetry records are hardly imaginable in production deployment. 

        The SPLT can be considered deployable only when the exporter limits the sequence to contain data about several first packets, usually in the order of tens of packets. Thus, the flow exporter ipfixprobe limits the size of the SPLT sequence to 30 packets. Furthermore, the Cisco Joy\footnote{\url{https://github.com/cisco/joy}} software can export SPLT sequences of up to 200 packets---with a default value of 50 packets. The short length then limits the SPLT applicability only to short connections.  Therefore, researchers proposed to extend the flows for additional raw features to capture information about the packets that do not fit into the sequence. For example, Tropkova et al.~\cite{sblt_our_paper} proposed to use a Sequence of Burst lengths and Times (SBLT), which carries the information about individual packet bursts (times and amount of transferred data). Nevertheless, even SBLT has also its length limit, and the aggregation of packets into the bursts loses some information about the exact timing of packets within the burst. 

        Extension for traffic histogram, which was used by Hofstede et al.~\cite{Hofstede2017} in the HTTPS brute force detection, also contains information about packets from the whole connection regardless of its length. However, similarly, as in SBLT, histograms do not carry information about the timing of individual packets; moreover, information about packet order is lost.

        Naturally, the previously described flow extensions can be combined. For example, Luxembruk et al.~\cite{LuxemburkQUIC} used SPLT and also traffic histograms in their QUIC classification task. Aceto et al.~\cite{ACETO2021102985} and Wang et al.~\cite{wang2017end} combine SPLT even with unencrypted information from packet headers. The combination with bytes of payload shows better performance. However, it also significantly increases network telemetry size by the first $x$ bytes of payload per flow. For example, both approaches use the first 784 bytes of transport-layer payload. When combined with SPLT of length 32, the size of each flow is 1296 bytes.

    \subsection{Extension for precomputed features}
        The extension for raw-packet information allows flexibility in feature extraction and allows each detector to compute its own specific discriminatory features; nevertheless, the flows with raw-packet information tend to be larger since they carry unprocessed raw data. Therefore, some flow exporters perform feature extraction during the flow creation process and provide extended flows with already extracted features. An example of such an exporter is the CICFlowMeter\footnote{\url{https://github.com/ahlashkari/CICFlowMeter}} that extends each flow with 80 statistical features---mainly mean, standard deviation, max, and min of multiple countable information from packets, such as the number of packets and TCP flags. These features are then used by multiple researchers in various network classification tasks~\cite{9796558,cic_bell_dns_2021_article,sharafaldin2018toward,Agrafiotis_Makri_Flionis_Lalas_Votis_Tzovaras_2022,ding2022imbalanced}.

        Similarly, as CICFlowMeter, MontazeriShatoori et al.~\cite{DoH_cic_dataset} created a DoHLyzer exporter\footnote{\url{https://github.com/ahlashkari/DoHLyzer}} that produces features directly within the flows, specifically for DoH detection task. These features are, however, entirely different from the CICFlowMeter---DoHLyzer and CICFlowMeter exported features are thus not universal. The inflexibility of the extensions for precomputed features then forces detectors to use non-optimal features, drastically degrading their performance. 

        The natural solution would be to export as many features as possible to cover a wide variety of detection/classification usecases. However, a large feature vector does not maintain the benefit of reduced flow size. For example, the list of discriminative features published by Moore et al.~\cite{Moore2005} in 2005 contains 249 different features, which would result in extended flows with unacceptable sizes of almost 1000\,bytes. Since 2005, various novel features have been used to target novel protocols and encrypted traffic analysis; thus, even a set of 249 features is by far not complete. 
        
        There are possible approaches for decreasing the size of large feature vectors, such as 1.\,compressive traffic analysis proposed by Nasr et al.~\cite{Nasr2017}, 2.\,use of embeddings created by Neural Networks instead of classical feature computation as proposed by Sungwoong et al.~\cite{Sungwoong2021}, 3.\,or feature dimension reduction techniques such as PCA (Principal Component Analysis)~\cite{yan2014}. However, we are not aware of any flow exporter that would deploy these techniques to reduce the feature set size and, thus the flow telemetry size. Moreover, all these approaches perform lossy compression that cannot be fully reconstructed back. The feature is then used in its compressed form, which is not understandable by people. Since people then analyze and inspect flows or alerts from the detectors, the lack of understandability then poses significant limitations in real-world deployment~\cite{Uhricek2023}.

        In our previous work~\cite{koumar2023}, we explored the universality features created by time-series analysis. The vector of 69 features showed excellent discriminative capabilities across multiple usecases; however, it was not deployable due to the computational complexity and memory requirements. Therefore, in this work, we focus on the flow extension of features based on time series analysis that are feasible to deploy in real-world monitoring infrastructure. The proposed feature vector contains features based on the statistical, time, distribution, and behavior properties acquired from the time series analysis of the SFTS. Furthermore, compared to all previous approaches, the universality and usefulness of the feature vector have been verified on 25 different network classification tasks using 15 network datasets.

\section{Datasets}
\label{sec:dataset_description}
    Since we aimed to find a universal feature vector for network traffic classification using machine learning, we needed to select multiple classification tasks that will be used for evaluation. We decided to select tasks based on evaluation data availability---the most studied and important network classification tasks have standard ``benchmark'' datasets that also allow comparison with the best-performing state-of-the-art classifiers. 

    We selected 15 different datasets that are provided in the form of raw packet captures (pcaps) and cover the most important detection and classification (even multiclass) tasks in network security. The used datasets for both binary and multiclass classification are written in~\tabref{tab:related_works_binary} along with the best-performing method. The table also contains the accuracies of the best-performing classifiers as well as their flow extension size.

    \begin{table*}[t!]
            \caption{Summarized related works for binary and multiclass classification. The Telemetry column represents the size of the flow extension (the classical flow has an extension size equal to zero). The Accuracy, F1-score, macro and weighted average F1-score are presented in percent [\%]. The ``mal.'' in column Task stands for malware, ``det.'' in column Task stands for detection, and ``class.'' stands for classification.}
            \begin{center}
                \begin{tabular}{lllrcc}
                    \toprule
                    \multicolumn{6}{c}{\textbf{Binary classification}} \\
                    \toprule
                    \textbf{Task} & \textbf{Dataset} &  \textbf{Approach} &  \textbf{Telemetry} &  \textbf{Accuracy} &  \textbf{F1-score} \\
                    \cmidrule{1-6}
                    Botnet det. & CTU-13 \cite{GARCIA2014100} & Stergiopoulos  et  al.   \cite{stergiopoulos2018automatic} & 1,000 & 99.85 & 99.90 \\
                    % detection & \multirow{-2}{*}{CTU-13 \cite{GARCIA2014100}} & Wang et al. \cite{wang2017malware}   & 784 & 99.41 & 99 \\ 
                    \hline
                    
                    Brute-force det. & HTTPS Brute-force \cite{jan_luxemburk_2020_4275775} & Luxemburk et al. \cite{9375998}  & 180 & 99.93 & 96.26  \\
                   
                    \hline
                    
                    Mining det. & CESNET-MINER22 \cite{richard_plny_2022_7189293} & Plný et al. \cite{plny2023decrypto} & 680 & 93.72 & 90.59  \\
                    \hline
                    
                    DNS malware det. & CIC-Bell-DNS \cite{cic_bell_dns_2021_article} & Kumaar et al. \cite{kumaar2021hybrid}  & 540 & 99.19 & 99.20  \\
                    % detection & \multirow{-2}{*}{CIC-Bell-DNS \cite{cic_bell_dns_2021_article}} & Mahdavifar  et  al.   \cite{cic_bell_dns_2021_article}  & 540 &	98.9 & 98.9  \\
                    
                    \hline
                    
                     & CIC-DoH-Brw \cite{montazerishatoori2020detection} &   Behnke et al. \cite{behnke2021feature} & 116 & -- & 99.8  \\
                    % \multirow{-2}{*}{DoH} & \multirow{-2}{*}{CIC-DoH-Brw \cite{montazerishatoori2020detection}} &  MontazeriShatoori et al. \cite{montazerishatoori2020detection} & 112  & -- & 99.30   \\
                   
                   \cline{2-6}
                    
                    \multirow{-2}{*}{DoH det.} & DoH-Real-World \cite{Jerabek2022} & Jeřábek et al.~\cite{doh_kamil} & 0 &  97.5 &  98.7 \\

                   \hline
                    
                    DoS attack det & Bot-IoT \cite{dos_iot_Dataset} &  Shafiq  et  al. \cite{SHAFIQ2020433} & 176 & 99.99 & 99.99 \\
                    % detection& \multirow{-2}{*}{Bot-IoT \cite{dos_iot_Dataset}} & Shafiq  et  al. \cite{SHAFIQ2020101863} & 40 & 99.99 & 99.99 \\
                    
                    \hline
                    
                     & IoT-23 \cite{sebastian_garcia_2020_4743746} &  Sahu et al. \cite{sahu2021internet}  & 144 & 96 &  96  \\
                     % & \multirow{-2}{*}{IoT-23 \cite{sebastian_garcia_2020_4743746}} &    Nascita et al. \cite{nascita2022machine}  & 11,520 & 99.93 & 91.70   \\
                    
                    \cline{2-6}
                   
                    IoT malware det. & Edge-IIoTset \cite{mbc1-1h68-22} &  Khacha et al. \cite{khacha2022hybrid}  & 472  & 99.99 &  99.99     \\
                    % classification & \multirow{-2}{*}{Edge-IIoTset \cite{mbc1-1h68-22}} & Ferrag et al. \cite{9751703}    & 72  & 99.99 & 99.96     \\
                    
                   \cline{2-6}
                    
                     & TON\_IoT \cite{moustafa2021new} &  Dai et al.  \cite{dai2023glads}  & 912 & 99.29  & 99.03 \\
                     % & \multirow{-2}{*}{TON\_IoT \cite{moustafa2021new}} & Guo \cite{guo2021machine}  & 36 & 99.23 & 98.90 \\
                   
                   \hline
                    
                     & CIC-IDS-2017 \cite{sharafaldin2018toward} &  Agrafiotis et al. \cite{Agrafiotis_Makri_Flionis_Lalas_Votis_Tzovaras_2022}  & 3,136 & 98.5 & 95.4  \\
                     % & \multirow{-2}{*}{CIC-IDS-2017 \cite{sharafaldin2018toward}} &    Ding et al. \cite{ding2022imbalanced}  & 316 & 95.86 &  95.81   \\
                    
                    \cline{2-6}
                   
                    \multirow{-2}{*}{Intrusion det.} & UNSW-NB15 \cite{moustafa2015unsw} & Nawir et al. \cite{nawir2018performances}  & 172 & 94.37 & 94.54  \\
                    % \multirow{-3}{*}{detection} & \multirow{-2}{*}{UNSW-NB15 \cite{moustafa2015unsw}} &   Mulyanto et al. \cite{mulyanto2020effectiveness}  & 472 & 86.73 &  90.41   \\
                   
                   \hline
                    
                    TOR det. & ISCX-Tor-2016 \cite{lashkari2017characterization} &  Dai et al.  \cite{dai2023glads}   & 912 & 99.99 & 99.65  \\
                    % detection & \multirow{-2}{*}{ISCX-Tor-2016 \cite{lashkari2017characterization}}  & Zhang et al. \cite{zhang2022identify}   & 270,000 & 99.99 & 99.01  \\
                    
                    \hline
                    
                      & ISCX-VPN-2016 \cite{icxs_vpn_2016_dataset} & Aceto et al. \cite{ACETO2021102985} & 1,296 & 93.75 & 91.95    \\
                    % & \multirow{-2}{*}{ISCX-VPN-2016 \cite{icxs_vpn_2016_dataset}} &  Shapira et al.~\cite{shapira2019flowpic} & 7.84\,MB & 88.4 & --  \\ 
                    
                    \cline{2-6}
                   
                    \multirow{-2}{*}{VPN det.} & VNAT \cite{vnat_dataset} & Jorgense et al.  \cite{vnat_dataset} & 3,612 & -- & 98.00  \\
                    % \multirow{-3}{*}{detection} & \multirow{-2}{*}{VNAT \cite{vnat_dataset}} &  Holodnak et al. \cite{10029386} & 2,000  &  -- & 85.30   \\
                    \bottomrule
                \end{tabular}
                \label{tab:related_works_binary}
            \end{center}
     % \end{table*}

     % \begin{table*}[t!]
     %        \caption{Summarized related works for multiclass classification. The Telemetry column represents the size of the flow extension (the classical flow has an extension size equal to zero).}
            \begin{center}
                \begin{tabular}{lllrccc}
                    \toprule
                    \multicolumn{7}{c}{\textbf{Mutliclass classification}} \\
                    \toprule
                     &    &  &  & & \textbf{Macro} & \textbf{Weighted} \\
                     \textbf{Task} & \textbf{Dataset} &  \textbf{Approach} &  \textbf{Telemetry} &  \textbf{Accuracy} &  \textbf{avg. F1} & \textbf{ avg. F1} \\

                    \cmidrule{1-7}

                    Botnet class. & CTU-13 \cite{GARCIA2014100} &  Marín et al. \cite{8844609} & 200 & 99.72 & 76.04 & 98.00 \\ 
                    % class. & \multirow{-2}{*}{CTU-13 \cite{GARCIA2014100}} &  Marín et al. \cite{8844609} & 200 & 99.72 & 76.04 & 98.00 \\ 

                    \hline
                    
                     IoT mal.  & Edge-IIoTset \cite{mbc1-1h68-22} & Khacha et al. \cite{khacha2022hybrid}  & 472 & 98.69 & -- & --   \\
                     % \multirow{-2}{*}{IoT } & \multirow{-2}{*}{Edge-IIoTset \cite{mbc1-1h68-22}} & Tareq et al.  \cite{tareq2022analysis}   & 2,832 & 94.94 & -- & -- \\
                    
                    \cline{2-7}
                    
                    class. & TON\_IoT \cite{moustafa2021new} & Tareq et al.  \cite{tareq2022analysis}   & 2,832 & 98.50 & 52.20 &  98.57  \\
                    % \multirow{-2}{*}{class.} & \multirow{-2}{*}{TON\_IoT \cite{moustafa2021new}}  &  Dai et al.  \cite{dai2023glads}   & 912 & 98.18  & 95.12 & -- \\
                   
                   \hline
                    
                     & CIC-IDS-2017 \cite{sharafaldin2018toward} &  Kunang et al. \cite{kunang2021attack}   & 784 &  95.79 & 84.54 & 95.11 \\
                    % & \multirow{-2}{*}{CIC-IDS-2017 \cite{sharafaldin2018toward}} & Leon et al.  \cite{9892293}   & 312 & 99.86 & -- & -- \\
                   
                    \cline{2-7}
                    
                    \multirow{-2}{*}{IDS class.} & UNSW-NB15 \cite{moustafa2015unsw} &   Madwanna et al.  \cite{madwanna2023yars}   & 472  & 82.21 &  53.15 & 80.30 \\
                    % \multirow{-3}{*}{class.} & \multirow{-2}{*}{UNSW-NB15 \cite{moustafa2015unsw}} & Mulyanto et al. \cite{mulyanto2020effectiveness}   & 14,160 & 73.39 & 39.78 & -- \\
                  
                    \hline  
                   
                    TOR class. & ISCX-Tor-2016 \cite{lashkari2017characterization} &  Dai et al.  \cite{dai2023glads}   & 912 & 97.95 & 86.77 & --  \\
                    % class. & \multirow{-2}{*}{ISCX-Tor-2016 \cite{lashkari2017characterization}} &  Shapira \cite{shapira2019flowpic} & 7.84MB  & 45.13 & 44.75 & 44.52 \\
                
                    \hline
                    
                     & ISCX-VPN-2016 \cite{icxs_vpn_2016_dataset} &   Dener et al.   \cite{dener2023rfse}  & 264 & 89.29 & 87.83 & 90.49 \\
                     % & \multirow{-2}{*}{ISCX-VPN-2016 \cite{icxs_vpn_2016_dataset}}  &  Aceto  et  al. \cite{ACETO2021102985}  & 1,296 & 73.14 & 71.14 & -- \\ 
                   
                    \cline{2-7}
                    
                    \multirow{-2}{*}{VPN class} & VNAT \cite{vnat_dataset} &  Jorgense et al.  \cite{vnat_dataset}  & 3,612 & 96 & -- & --  \\
                    % \multirow{-3}{*}{class.} & \multirow{-2}{*}{VNAT \cite{vnat_dataset}} &  Holodnak et al. \cite{10029386}  & 2,000 & -- & 85.30 &  -- \\
                    \bottomrule
                \end{tabular}
                \label{tab:related_works_multiclass}
            \end{center}
     \end{table*}

    To find the best-performing classifiers for each dataset, we used common scientific manuscript aggregators such as IEEE Explore\footnote{\url{https://ieeexplore.ieee.org/}}, ACM Digital Library\footnote{\url{https://dl.acm.org}}, Scopus\footnote{\url{https://www.scopus.com}}, and Google Scholar\footnote{\url{https://scholar.google.com}}. We either listed articles that referenced concerned datasets or searched for the names of datasets. We went through more than 300 papers and selected the best-performing proposals that met the following conditions ensuring fair performance comparability with related works: 1.~it was a flow-based method, 2.~it used the dataset as a whole and classifies all the dataset classes and types of samples, 3.~did not use IP addresses and transport layer ports as input features\footnote{The concerned datasets are mainly lab-created; thus usage of IP addresses and transport layer ports is not considered---in this case---as a good practice due to dataset overfitting as described by Behnke et al.~\cite{behnke2021feature}}, 4.~did not combine the concerned dataset with additional data. The selected methods are then used for performance and telemetry-size comparison with the novel feature vector.

\section{Time Series Analysis} \label{tsa_in_flow_exporter}

    The time series is a natural representation of network packets transferred via the computer networks. The SPLT is an example of time series, where each packet and its timestamp represent a single point. Nevertheless, SPLTs are usually very short, containing about tens of data points. The ipfixprobe flow exporter exports SPLT of length 30, which is sometimes insufficient. Even though the majority of flows on the internet are short---so-called dragonfly flows~\cite{brownlee2002understanding}, these statistics are often highly influenced by DNS flows with only two packets. According to Luxemburk et al.~\cite{luxemburk2022fine}, more than 15\% of TLS flows with successful TLS handshake on the internet are longer than 30 packets, and in the case of QUIC flows, more than 40\% of flows are longer than 30 packets~\cite{luxemburk2023cesnet}. The SPLT sequence then loses information for a non-negligible portion of traffic that could be otherwise used for more accurate detection. 
    
\begin{figure*}
            \centerline{\includegraphics[width=15cm]{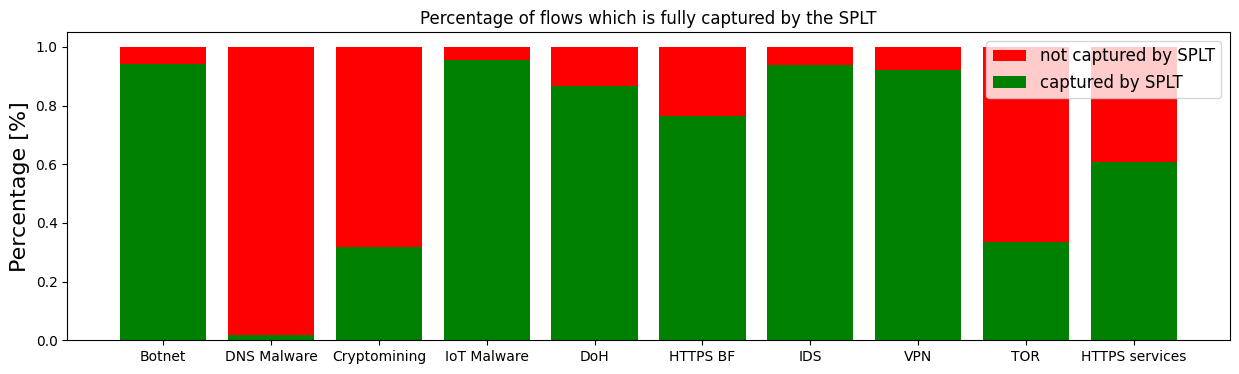}}
            \caption{The shares of network flows in selected datasets that can (cannot) be captured by SPLT of length 30}
            \label{fig:percentage_splt_vs_nettisa}
        \end{figure*}

    The information loss is also demonstrated in~\figref{fig:percentage_splt_vs_nettisa}, which shows the percentage of flows that contain more than 30 packets in each selected detection task in~\secref{sec:dataset_description}. It can be seen that the flow length depends on the classification tasks. In the case of TOR, Cryptomining, and DNS malware, only a small portion of traffic is captured by SPLT. Nevertheless, all classification tasks would benefit from features that can capture flows longer than 30 packets.

    The benefit of a longer SPLT sequence can be seen in~\figref{fig:percentage_splt_vs_nettisa}, where each traffic type generates different patterns of packet sizes in time. These patterns can be captured by time series analysis~\cite{KoumarUSTS}; thus, features based on time series analysis are an ideal candidate for universal feature vectors as previously published in our study~\cite{koumar2023}. Nevertheless, the previous features were computationally and memory intensive, which prevented their deployment in high-speed networks. Therefore, we further explored the possibility of lightweight time series analysis that would allow deployment in all network infrastructures, even the large and high-speed ones.

        \begin{figure*}[ht!]
                \centerline{\includegraphics[width=17cm]{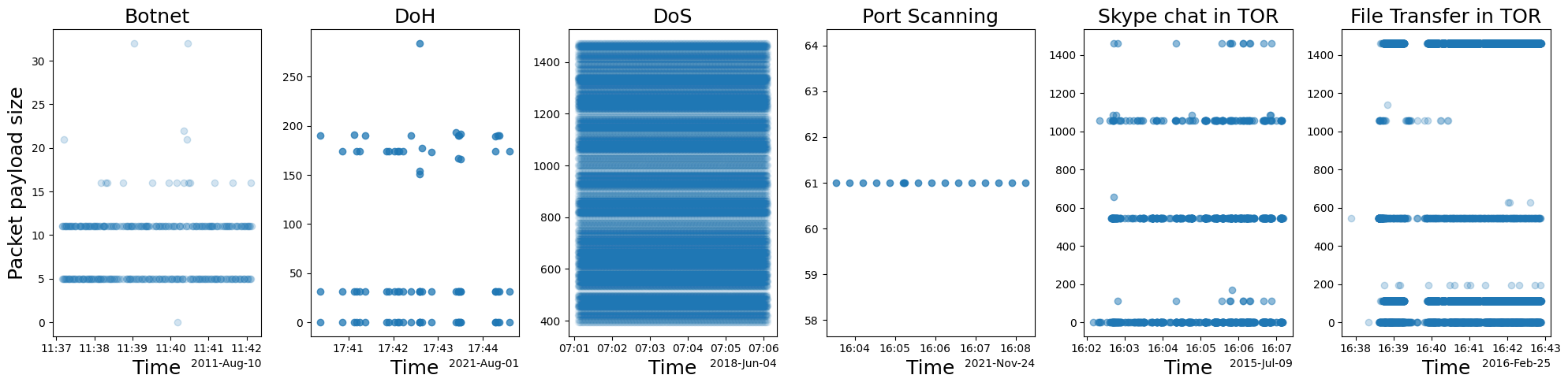}}
                \caption{Examples of packet time series of different traffic types.}
                \label{plot:sfts_examples_behaviors}
        \end{figure*}

    \subsection{Background in Time Series Analysis}
        The state of the art in time series analysis mostly considers only evenly-spaced times between observations. This type of time series is called evenly spaced (or regularly sampled) \cite{hamilton2020time}, and it is defined as the sequence of observation \(\{X_n\} = \{x_1, \dots, x_n\}\) taken in times \(\{T_n\} = \{t_1, \dots, t_n\}\), where \(n\) is the number of observations. It is always true that \(t_{j+1} - t_{j} = t_{j} - t_{j-1}, \forall j \in \{2, \dots, n\}\). Because of this behavior, it is possible to apply division and get the sequence of times \(\{T_n\} = \{1,2, \dots, n\}\). So when an evenly spaced time series is used, then it is written only as \(\{X_n\}\) where \(n := 1,2,\dots,n\) and absolute observation times are unnecessary.  

        The evenly-spaced time series are often used in network traffic analysis, mainly for forecasting and anomaly detection~\cite{moayedi2008arima,cook2019anomaly}. 
        Some previous works~\cite{montazerishatoori2020detection} use evenly spaced time series even for classification. However, network traffic naturally occurs with unevenly spaced timestamps (packet transmission time). 
        
        To create an evenly spaced time series from network traffic, we need to set the aggregation interval---the time window for a single datapoint in the series---that highly affects the analysis result due to packets occurring at the aggregation interval borders. Moreover, a majority of evenly-spaced time series from network traffic contains one or more intervals where no packet was transmitted---zero-value intervals. Badly selected aggregation intervals cause analysis failure: A large aggregation window causes a loss of information in behavior patterns; a small aggregation window results in a large number of zero values, which highly affects the analysis results. Unfortunately, each time series has a different ideal aggregation interval, which can also be changed in time---thus, the analysis failure with evenly spaced time series is inevitable~\cite{KoumarUSTS}. 
        
        In our approach, we create a time series from packets within a flow---the series of payload sizes in bytes with the corresponding transmission timestamp. We call them Single Flow Time Series (SFTS). However, the SFTS created by the sizes of packets and their timestamps do not have evenly spaced timestamps between the datapoints. That means a time series of observations \( \{X_n\} = \{x_1, x_2, \dots, x_n\}\) taken at times \( \{T_n\} = \{t_1, t_2, \dots, t_n\}\) does not have constant \(\delta_j = t_{j+1} - t_{j}, \forall j \in \{1,\dots, n-1\}\). This type of time series is called unevenly (or unequally/ irregularly) spaced.

        The unevenly-spaced time series from network traffic does not require any aggregation; moreover, they do not contain any trends and sesonality~\cite{KoumarUSTS}, which makes them more suitable for network traffic classification. Since there is already an advanced mathematical theory around unevenly-spaced time series (mainly in astronomy and medicine), we can use the mathematical methods and tools for their analysis to form a feature vector describing the properties of the communication and use it for network traffic classification~\cite{koumar2023}. Nevertheless, we also need to keep the computational and memory complexity in mind when designing the feature vector, to ensure the possibility of practical deployment.

\section{Features description} \label{exported_features}
    This section contains a detailed description of individual features that are exported in the novel extended flow---NetTiSA flow. We divide the NetTiSA classification features into three groups by computation. The first group of features is based on classical bidirectional flow information---a number of transferred bytes, and packets.  The second group contains statistical and time-based features calculated using the time-series analysis of the packet sequences. The third type of features can be computed from the previous groups (i.e., on the flow collector) and improve the classification performance without any impact on the telemetry bandwidth. 
    
    \subsection{Flow features}
        In the flow features, we explicitly omit the use of IP addresses and transport ports, since including them in the feature vectors is generally considered a bad practice, as described by Behnke et al.~\cite{behnke2021feature}. The flow features are:
            % [style=multiline,leftmargin=4cm,font={\itshape}]
            \begin{description}
                \item[Packets] is the number of packets in the direction from the source to the destination IP address.
                \item[Packets in reverse direction] is the number of packets in the direction from the destination to the source IP address.
                \item[Bytes] is the payload size in bytes transferred in the direction from the source to the destination IP address.
                \item[Bytes in reverse direction] is the payload size in bytes transferred in the direction from the destination to the source IP address.
            \end{description}

\subsection{Statistical and Time-based features}
    These features are exported in the extended part of the flow. All of them can be computed (exactly or in approximative) by streamwise computation, which is necessary for keeping memory requirements low. It contains the following features:

    % [style=multiline,leftmargin=4cm,font={\itshape}
    \begin{description}
        \item[Mean] represents mean of the payload lengths of packets with equation \( \mu =  \frac{1}{n} \sum_{i=1}^{n} x_{i}\). 
    
        \item[Min] is the minimal value from payload lengths of all packets in a flow. 
        % It requires only storing one integer and comparison when a new packet arrives.
        \item[Max] is the maximum value from payload lengths of all packets in a flow. 
        % It requires only storing one integer and comparison when a new packet arrives.
        \item[Standard deviation] is a measure of the variation of payload lengths from the mean payload length. In general it can be computed with equation \(\sigma = \sqrt{\frac{1}{n} \sum_{i=1}^{n} \left(x_{i} - \mu\right)^{2} } \). However, the mean value \(\mu\) is not available for streamwise computation. Therefore, we use the existing streamwise equation \(\sigma = \sqrt{ \frac{\sum_{i=1}^{n} x_{i}^{2}}{n} - \left( \frac{\sum_{i=1}^{n} x_{i}}{n} \right)^{2} } \).      
        \item[Root mean square] is the measure of the magnitude of payload lengths of packets computed as  \( rms = \sqrt{ \frac{1}{n} \sum_{i=1}^{n} x_{i}^{2} } \). 
        
        % It requires only two additional operations before exporting because all other is done for the \textit{Standard deviation} features.
        \item[Average dispersion] is the average absolute difference between each payload length of the packet and the mean value computed as \( ad = \frac{1}{n} \sum_{i=1}^{n} \left|x_{i} - \mu\right| \).  For the purpose of streamwise computation, it must be computed with approximated mean  \( ad = \frac{1}{n} \sum_{i=1}^{n} \left|x_{i} - \hat{\mu}_i\right| \) which is computed with equation \(\hat{\mu}_i = \hat{\mu}_{i-1} +  \frac{x_{i} - \hat{\mu}_{i-1}}{i}\), where \(i := 1,2,\dots\) and \(\hat{\mu}_0 := 0\). This equation allows computation in a streamwise manner, without the need to store the whole payload sequence.
        
        % Furthermore, computation of approximated mean cost one addition, substitution, and division operation  when a new packet arrives. The average dispersion sum requires one addition and substitution operation when a new packet arrives. 
        \item[Kurtosis] is the measure describing the extent to which the tails of a distribution differ from the tails of a normal distribution. It is computed by \( Kurt = \frac{1}{n \sigma^{4}} \sum_{i=1}^{n} \left(x_{i} - \mu\right)^{4} \). For streamwise computation, we will use the approximated mean as for the feature \textit{Average dispersion}.
        
        % It requires one additional integer to store, operation of amplification, and additional operation. The \(x_{i} - \hat{\mu}_i\) is already computed for the \textit{Average dispersion}.
        \item[Mean of relative times] is the mean of the relative times which is a sequence defined as \(\{st\} = \{ t_1 - t_1, t_2 - t_1, \dots, t_n - t_1 \}\). 
        % We compute the mean of the value with the same method as for feature \textit{Mean}.
        \item[Mean of time differences] is the mean of the time differences which is a sequence defined as \newline \(\{dt\} = \{ t_j - t_i | j = i + 1, i \in \{1, 2, \dots, n - 1\} \}\). 
        % We compute the mean of the value with the same method as for feature \textit{Mean}.
        \item[Min from time differences] is the minimal value from all time differences, i.e., min space between packets.
        \item[Max from time differences] is the maximum value from all time differences, i.e., max space between packets.
        \item[Time distribution] describes the deviation of time differences between individual packets within the time series. The feature is computed by using the following equation:
         
         \begin{equation}
            tdist = \frac{ \frac{1}{n-1} \sum_{i=1}^{n-1} \left| \mu_{\{dt_{n-1}\}} - dt_i \right| }{ \frac{1}{2} \left(max\left(\{dt_{n-1}\}\right) - min\left(\{dt_{n-1}\}\right) \right) }     
         \end{equation}

        The maximum and minimum of the time differences are also calculated as a feature and the mean of the time differences is approximated the same way as for the mean calculated for the purpose of the \textit{Average dispersion} and \textit{Kurtosis} features. The lower the \(tdist\), the better the time differences spread over time. 
    
        \item[Switching ratio] represents a value change ratio (switching) between payload lengths. The switching ratio is computed by 
            $ sr = \frac{s_n}{\frac{1}{2} (n - 1)} $
        where \(s_n\) is number of switches. Its streamwise computation requires storing two integer variables only. One integer is for counting a number of different payload lengths of consecutive packets, and the second integer is for storing the last payload length value.
    \end{description}

    \subsection{Features computed at the collector}
    The third set contains features that are computed from the previous two groups before the classification itself. Therefore, they do not influence the network telemetry size and their computation does not put additional load to resource-constrained flow monitoring probes. The \textit{NetTiSA flow} combined with this feature set is called the \textit{Enhanced NetTiSA flow} and contains the following features: 

    % [style=multiline,leftmargin=4cm,font={\itshape}]
    \begin{description}
        \item[Max minus min]  is the difference between minimum and maximum payload lengths. 
        \item[Percent deviation] is the dispersion of the average absolute difference to the mean value computed as \( pd = \frac{ad}{\mu} \).
        \item[Variance] is the spread measure of the data from its mean calculated as the square root of the \textit{Standard deviation}.
        \item[Burstiness] is the degree of peakedness in the central part of the distribution computed as \( b_{x_{n}} = \frac{\sigma - \mu}{\sigma + \mu}\).
        \item[Coefficient of variation] is a dimensionless quantity that compares the dispersion of a time series to its mean value and is often used to compare the variability of different time series that have different units of measurement. It is calculated as \( cv = \frac{\sigma}{\mu} \). 
        \item[Directions] describe a percentage ratio of packet direction computed as \(\frac{d_1}{ d_1 + d_0}\), where \(d_1\) is a number of packets in a direction from source to destination IP address and \(d_0\) the opposite direction. Both  \(d_1\) and \(d_0\) are inside the classical bidirectional flow. 
        
        If they are all in the direction from the source to the destination, then the percentage is 100\%; if they are all from the destination to the source, then the percentage is 0\%.
        \item[Duration] is the duration of the flow.
    \end{description}

    \subsection{Feature extraction}
        We extracted flows with all proposed features from every PCAP from datasets listed in~\tabref{tab:related_works_binary}. The flow extraction process was set with 300 seconds of active timeout and 65 seconds of inactive timeout\footnote{If no packet is observed within the ``inactive timeout'' period, the flow is considered terminated. Flows longer than the ``active timeout'' are split and are exported every time this timeout elapses.}. Moreover, when a feature could not be computed (such as Standard deviation, or Switching ratio) due to a lack of data in short single-packet flows, we filled its value with ``0''. Finally, we split the datasets\footnote{We used stratified sampling so that the original ratio of label classes remains in the split parts.} of flow data extended for Enhanced NetTiSA features into a standard train, validation, and test parts and their detailed description is shown in~\tabref{tab:datasets_tsa_sfts}. We made the created datasets publicly available using the Zenodo platform\cite{OUR_DATASET} to facilitate reproducibility.

        \begin{table*}[t!]
            \caption{Description of NetTiSA flow datasets created from original datasets.}
            \begin{center}
                \begin{tabular}{QN{3.9cm}O{0.8cm}O{0.8cm}O{1.8cm}O{1.8cm}O{1.8cm}}
                    \toprule
                    \multicolumn{7}{c}{\textbf{Binary classification}} \\
                    \toprule
                     \textbf{} &  &  \multicolumn{2}{c}{\textbf{Labels [\%]}}  &   \multicolumn{3}{c}{\textbf{Sizes of datasets}}  \\
                    \cmidrule{3-7}
                     \textbf{Task} & \textbf{Original dataset} & \textbf{False} & \textbf{True} & \textbf{Train} & \textbf{Validation} & \textbf{Test} \\
                    % \hline
                    \toprule
                    Botnet det. & CTU-13 \cite{GARCIA2014100}  & 88.57 & 11.43 & 194,928 & 83,540  & 50,000 \\
                    \hline
                    Cryptomining det.  & CESNET-MINER22 \cite{richard_plny_2022_7189293} &  64.97 & 35.03  & 1,417,433 & 607,470 & 1,075,576 \\
                    \hline
                    DNS Malware det. & CIC-Bell-DNS-2021 \cite{cic_bell_dns_2021_article}  &  96.04 & 3.96  &  3,761 & 1,613  & 1,000  \\
                    \hline
                     & CIC-DoHBrw-2020 \cite{montazerishatoori2020detection}  & 69.48 & 30.52 & 623,952 & 267,408 & 100,000 \\
                    \multirow{-2}{*}{DoH det.} & DoH Real-world \cite{Jerabek2022}  & 1.52 & 98.47  & 3,564,331 & 1,527,571 & 500,000 \\
                    \hline
                    DoS det. & Bot-IoT \cite{dos_iot_Dataset}  &  97.40 & 2.59  & 2,106,874  & 902,946 & 1,000,000  \\
                    \hline
                    Brute-force det. & HTTPS Brute-force \cite{jan_luxemburk_2020_4275775} &  95.71 & 4.29  & 646,366 & 277,014 & 100,000  \\   
                    \hline
                     & Edge-IIoTset \cite{mbc1-1h68-22}  & 92,95 & 7,05 & 827,557  & 354,668 &   250,000  \\
                     & IoT-23 \cite{sebastian_garcia_2020_4743746} &  24.56 & 75.44 &  2,492,447 & 1,068,192 &  500,000  \\
                    \multirow{-3}{*}{IoT Malware det.} & TON\_IoT \cite{moustafa2021new} & 0.31 & 99.69 &  2,077,190 & 890,224 & 500,000  \\
                    \hline
                      & CIC-IDS-2017 \cite{sharafaldin2018toward} &   77.60 & 22.40 & 1,182,652 & 506,850 &  500,000 \\
                    \multirow{-2}{*}{Intrusion det.} & UNSW-NB15 \cite{moustafa2015unsw} & 96.50  & 3.50 & 954,370 & 409,016 & 584,309  \\
                    \hline
                     TOR det. &  ISCX-Tor-2016 \cite{lashkari2017characterization}  &  98.99 & 1.01  & 19,200 &  8,229 & 11,756  \\
                    \hline
                     &  ISCX-VPN-2016 \cite{icxs_vpn_2016_dataset}  &  90.54 & 9.46  &  135,715 & 58,164 & 50,000 \\
                    \multirow{-2}{*}{VPN det.}  & VNAT \cite{vnat_dataset} &  96.61 & 3.39  & 23,380 & 10,020 & 10,000  \\
                    \bottomrule
                \end{tabular}
                %\label{tab:datasets_tsa_sfts}
            \end{center}

            \begin{center}
                \begin{tabular}{QN{6.1cm}O{1.8cm}O{1.8cm}O{1.8cm}}
                    \toprule
                    \multicolumn{5}{c}{\textbf{Multiclass classification}} \\
                    \toprule
                     \textbf{} &  &   \multicolumn{3}{c}{\textbf{Sizes of datasets}}  \\
                    \cmidrule{3-5}
                     \textbf{Task} & \textbf{Original dataset} & \textbf{Train} & \textbf{Validation} & \textbf{Test} \\
                    \toprule
                    All binary class.  &  & 10,934,721 & 4,686,309 &  4,000,000  \\
                    \cline{1-1} \cline{3-5}
                    All multiclass class.  & \multirow{-2}{6.1cm}{\cite{GARCIA2014100,richard_plny_2022_7189293,cic_bell_dns_2021_article,montazerishatoori2020detection,Jerabek2022,dos_iot_Dataset,jan_luxemburk_2020_4275775,mbc1-1h68-22,sebastian_garcia_2020_4743746,moustafa2021new,sharafaldin2018toward,moustafa2015unsw,lashkari2017characterization,icxs_vpn_2016_dataset,vnat_dataset}}   & 7,127,894 & 3,054,812 &  9,999,803  \\
                    \hline
                     Botnet class. & CTU-13 \cite{GARCIA2014100}  & 106,212 & 45,519  & 25,000   \\
                    \hline
                     &  CIC-IDS-2017 \cite{sharafaldin2018toward}  & 1,392,651 & 596,851 & 200,000  \\
                    \multirow{-2}{*}{Intrusion  class.} &   UNSW-NB15 \cite{moustafa2015unsw} & 954,370 & 409,016 & 584,309  \\
                    \hline
                     & Edge-IIoTset \cite{mbc1-1h68-22}  & 827,557  & 354,668 &   250,000  \\
                     \multirow{-2}{*}{IoT Malware class.}  &  TON\_IoT \cite{moustafa2021new}  & 2,252,190 &  965,224 &  250,000  \\
                    \hline
                     TOR class. &  ISCX-Tor-2016 \cite{lashkari2017characterization}    & 70,557 & 30,239  &  20,000  \\
                    \hline
                      &  ISCX-VPN-2016 \cite{icxs_vpn_2016_dataset}  & 12,655 & 5,423  &  5,000  \\
                    \multirow{-2}{*}{VPN class.}  & VNAT \cite{vnat_dataset}   & 23,380 & 10,020  & 10,000  \\
                    \bottomrule
                \end{tabular}
                \label{tab:datasets_tsa_sfts}
            \end{center}
        \end{table*}

\section{Feature Discriminative Performance} 
\label{classification_section}
        The discriminative performance of features was evaluated by using selected datasets (described in~\secref{sec:dataset_description}) and appropriate network traffic classification tasks as listed in~\tabref{tab:related_works_binary}. We created a novel traffic classifier for each task and evaluated its performance. The performance of the classifiers then serves as a Feature Discrimination measure.
        
        The classifier creation pipeline consisted of three steps. At first, we select the optimal ML algorithms. We tested 11 well-known ML algorithms, including kNN, SVM, Random Forest, XGBoost, and AdaBoost, from which the XGBoost algorithm outperforms all others on all classification tasks. 
        
        As a second step, we tuned the XGBoost hyperparameters of the ML algorithms for each dataset and classification task using the validation dataset to avoid hyperparameter overfitting~\cite{ghojogh2019theory}. We use the \textit{hyperopt library}~\cite{bergstra2013making} for tuning the following hyper-parameters: \textit{n\_estimators}, \textit{max\_depth}, \textit{gamma}, \textit{reg\_alpha}, \textit{reg\_lambda}, \textit{min\_child\_weight}, and \textit{colsample\_bytree}.
        The rest of the hyper-parameters were left at default values.  
        % are shown in the~\ref{sec:hyperparameters} of the paper in~\tabref{tab:hyperparameters}. %Appendix is in the reference name

        % \begin{figure}[h]
        %     \centerline{\includegraphics[width=9cm]{pictures/algorithm_selection_2.png}}
        %     \caption{Results of Machine Learning algorithm selection on the problem of detection of the Botnet.}
        %     \label{fig:algorithm_selection}
        % \end{figure}

        As a third step, we trained the best-performing XGBoost model with previously obtained hyperparameter values. The trained models were then evaluated on the test part of the dataset using \textit{Accuracy} and \textit{F1-score} classification metrics, which are defined in the following formulas:
        
        % % \begin{equation}
        %     \begin{gather} 
        %         Accuracy = \frac{TP + TN}{TP + TN + FP + FN} \\ 
        %         Precision = \frac{TP}{TP + FP} \\
        %         Recall = \frac{TP}{TP + FN} \\
        %         F1-score = \frac{2 * Precision * Recall}{Precision + Recall}
        %     \end{gather}
        % % \end{equation}

        \begin{equation}
                Accuracy = \frac{TP + TN}{TP + TN + FP + FN} \\ 
        \end{equation}
        \begin{equation}
                Precision = \frac{TP}{TP + FP} \\
        \end{equation}
        \begin{equation}
                Recall = \frac{TP}{TP + FN} \\
        \end{equation}
        \begin{equation}
                F1-score = \frac{2 * Precision * Recall}{Precision + Recall}
        \end{equation}
        
        Where \(TP\) is number of \textit{True positive}, \(TN\) is number of \textit{True negative}, \(FP\) is number of \textit{False positive}, and \(FN\) is number of \textit{False negative}. Furthermore, we use macro and weighted averages of these metrics for multiclass classification. The macro average is the arithmetic mean, and the weighted average considers the support of each class.

        The test part was not used during any stage of the classifier design, ensuring the fairness of model evaluation on data that was not seen before. The source codes of our whole classification pipeline, including the pre-processing and the final settings of the hyperparameters for each classification problem, are publicly available in our repository\footnote{\url{https://github.com/koumajos/Classification_by_NetTiSA_flow}}.

    \subsection{Classification results} \label{sec:classification_results}
        In this section, we present classification performance on 25 selected datasets. The performance of the classifier trained on the novel Enhanced NetTiSA features is always compared with the best-performing classifier from the related work. Moreover, we also compare the performance with our previously published results~\cite{koumar2023}, since this is the only work that uses Time Series Analysis of Single Flow Time Series.

        \subsubsection{Binary classification}
            The results of binary classification are graphically presented in \figref{fig:results_binary_classification}. We also provide detailed results in Appendix, in Tables \ref{tab:classification_results_binary} and \ref{tab:confusion_binary}. 

            \begin{figure*}[t!]
                \centerline{\includegraphics[width=16cm]{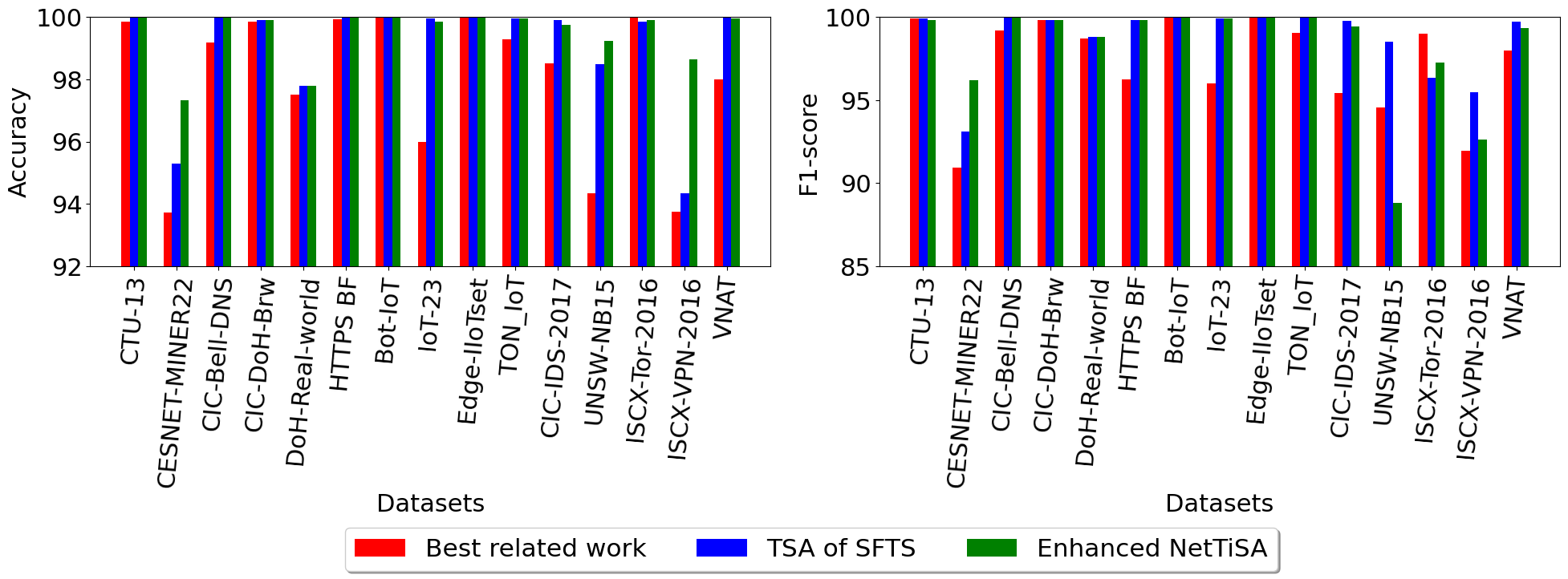}}
                \caption{Comparison of binary classification by best-related works, complete feature set based on TSA of SFTS and Enhanced NetTiSA flow. We compare approaches by Accuracy and F1-score (in [\%]).}
                \label{fig:results_binary_classification}
            \end{figure*}

            The classification based on the Enhanced NetTiSA feature set achieved mostly similar or better results than the best-performing classifier from related works\footnote{The best-performing related works are written for each dataset in \tabref{tab:related_works_binary}}. On six binary classification tasks, the Enhanced NetTiSA features performed significantly better---it achieved more than 1\% increase in accuracy measure or F1 score compared to related works. Nevertheless, with the TOR detection problem Intrusion detection on the UNSW-NB-15 dataset, we observed a slightly worse F1 score than the best-performing classifier from related work. 

            We investigated related works of Intrusion detection on the UNSW-NB-15 dataset. The classifier published by Nawir et al.~\cite{nawir2018performances} performed significantly better in F1-score (by 5.75\%). However, their accuracy is significantly worse than in the case of Enhanced NetTiSA (by 4.85\%). The increased F1 score can be attributed to the application-layer-based features (such as from HTTP headers) that would not be applicable in encrypted traffic analysis. 

            We also investigated the difference between the TOR classifier published by Dai et al.~\cite{dai2023glads}, which extends flow data for the first 32 packets and also 600 bytes of payload. The payload information gives their classifier an advantage since the classifier can fit the unencrypted TLS handshake data. Since the TOR usually does not use some common TLS extensions, e.g., Server Name Indication, the TOR TLS handshakes in the dataset are specific and easily recognizable by payload-based analysis. Nevertheless, the raw payload data gives Dai et al. an advantage, resulting in also in larger telemetry records as seen in \figref{fig:telemetry_comparsion}. 

            When we compare the Enhanced NetTiSA discriminative performance with universal Time-Series Analysis features proposed by Koumar et al.~\cite{koumar2023}, we can see (in \figref{fig:results_binary_classification}) that the results are mostly similar. The features proposed by Koumar et al. show significantly better discriminative performance (when concerning the F1-score) in the ISCX-VPN-2016 and  UNSW-NB-15 datasets. Nevertheless, the NetTiSA significantly outperformed Koumar et al.~\cite{koumar2023} on the CESNET-MINER22 dataset. The overall comparison of the binary classification is summarized in~\tabref{tab:tsa_sfts_vs_nettisa}.
            
            %On the binary classification, the NetTiSA achieves similar results in most of the classification tasks as can be seen in Figure \ref{fig:results_binary_classification}. Furthermore, on the cryptomining detection, the NetTiSA significantly outperforms the TSA of SFTS by more than 2\% Accuracy and 3\% F1-score. However, on problems of the VPN detection using the ISCX-VPN-2016 dataset, Tor detection using ISCX-Tor-2016 dataset, and Intrusion detection using the UNSW-NB-15 dataset, although the NetTiSA achieved a significantly higher accuracy, it also achieved a significantly lower F1-score. In each case, the TSA of SFTS feature vector contains features that in a lot of cases end with NaN value (mainly frequency and distribution-based features). So, the deletion of rows with NaN values results in different dataset sizes and label ratios. Therefore, the lower F1-score of NetTiSA is mainly caused by an imbalance of False and True classes.

        \subsubsection{Multiclass classification}

            The results of multiclass classification are presented in~\figref{fig:results_multiclass_classification}. Furthermore, the detailed results are also presented in the Appendix, Tables~\ref{tab:classification_results_mutliclass}, \ref{tab:confusion_ctu_13}, \ref{tab:confusion_vpn_iscx}, \ref{tab:confusion_vpn_vnat}, \ref{tab:confusion_edge_iiot}, \ref{tab:confusion_ton_iot}, \ref{tab:confusion_ids_cic}, \ref{tab:confusion_ids_unsw}, and \ref{tab:confusion_tor}. Similarly, as in binary classification, the Enhanced NetTiSA features proved to be discriminatory in the multiclass classification. It outperformed most of the best-performing classifiers proposed in related works (written in~\tabref{tab:related_works_multiclass}). Specifically, in five out of eight cases, we achieved more than a 1\% classification performance increase compared to best-performing related works. However, we observed a slight decrease in two cases---TON\_IoT and TOR cases. 

            \begin{figure*}[t!]
                     \centerline{\includegraphics[width=16cm]{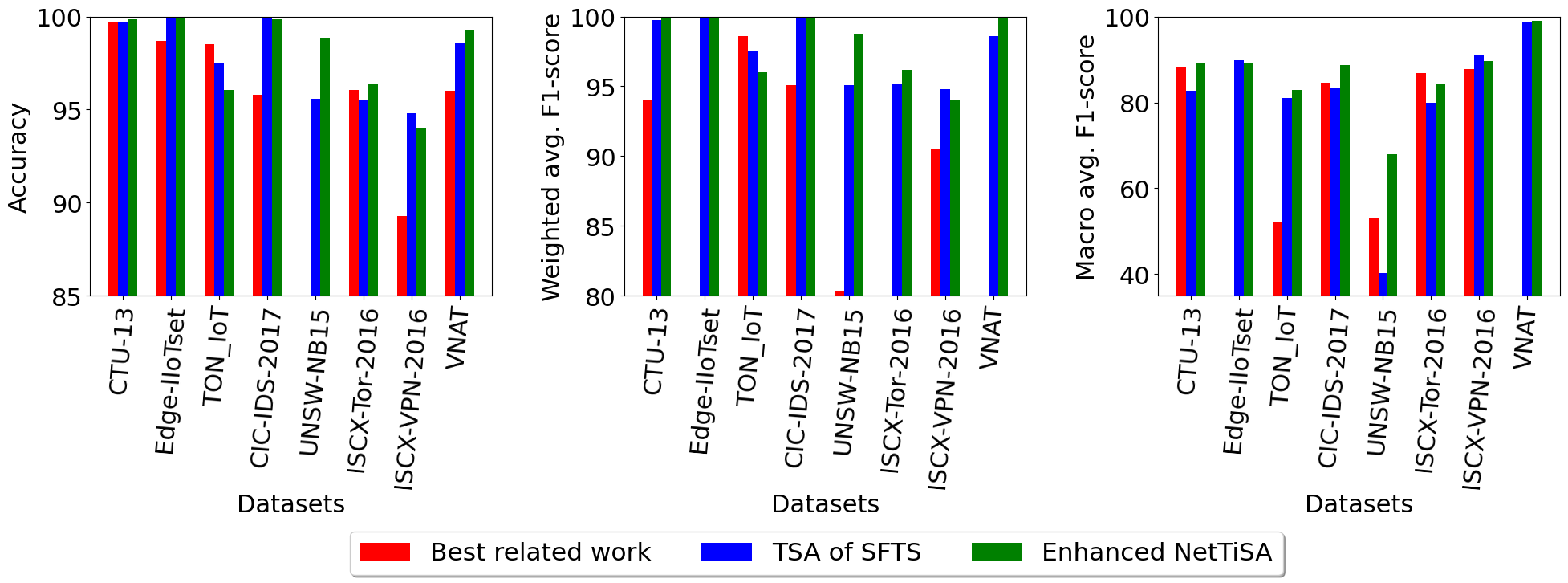}}
                    \caption{Comparison of multiclass classification by best-related works, complete feature set based on TSA of SFTS and Enhanced NetTiSA flow. We compare approaches by Accuracy, macro and weighted average F1-score (in \%).}
                    \label{fig:results_multiclass_classification}
            \end{figure*}

            The best-performing classifier of TON\_IoT published by Tareq et al.~\cite{tareq2022analysis} is based on a 2D convolutional network (CNN) with very long packet-length data (SPLT with all packets from connection) organized in the image. The SPLT data give the classifier an advantage in the opportunity of high-quality feature extraction that allows accurate classification. Nevertheless, the long packet sequences (SPLT) cannot be exported in real-world deployment scenarios due to the technical limitations of the flow exporters (see~\secref{related_works_section}). Furthermore, although the Tareq et al.~\cite{tareq2022analysis} achieves slightly better accuracy and weighted average F1-score (by almost 2.5\%), they only achieved 52.2\% macro average F1-score, which is 30.62\% less than the classifier that uses Enhanced NetTiSA feature set. 

            Similarly, as in the binary classification of TOR, Dai et al.~\cite{dai2023glads} achieved better Macro average F1-score with TOR of 86.77\%, while the classifier trained with NetTiSA achieved only 79.87\%. Nevertheless, the raw payload data results in larger telemetry records, as seen in \figref{fig:telemetry_comparsion}. 

            When we compare the Enhanced NetTiSA with the only universal feature set proposed by Koumar et al.~\cite{koumar2023},  we can see that the novel Enhanced NetTiSA performs significantly better in four out of eight evaluated cases. In only one case---TON\_IoT---the novel Enhanced NetTiSA features performed significantly worse (approx 1.5\% lower Accuracy Weighted average F1-score) than the Koumar et al.~\cite{koumar2023}. The overall comparison with Koumar et al.~\cite{koumar2023} is shown in~\tabref{tab:tsa_sfts_vs_nettisa}. It can be seen, that on multiclass classification, the NetTiSA performs on average better in all performance metrics.

            \begin{table*}[t!]
                \caption{Comparison of average results of binary and multiclass classification based on NetTiSA flows with classification based on all features generated using TSA of SFTS. Comparison is made by the average and standard deviation of each classification metric across all datasets (in \%).}
                \begin{center}
                    \begin{tabular}{l c c|c c c}
                        \toprule
                        & \multicolumn{2}{c|}{\textbf{Binary classification}}  & \multicolumn{3}{c}{\textbf{Multiclass classification}} \\
                        \toprule
                        &  &  &  & \textbf{Weighted avg.} & \textbf{Macro avg.} \\
                        & \textbf{Accuracy} & \textbf{F1-score} & \textbf{Accuracy} & \textbf{F1-score} & \textbf{F1-score} \\
                        \toprule
                        \textbf{NetTiSA} & 99.48 ($\pm$0.840) & 98.13 ($\pm$3.167) & 98.03 ($\pm$2.098) & 86.34 ($\pm$8.267) & 98.06 ($\pm$2.186) \\
                        \textbf{TSA of SFTS}  & 99.02 ($\pm$1.772) & 98.74 ($\pm$2.026) & 97.70 ($\pm$2.025) & 80.87 ($\pm$16.480) & 97.59 ($\pm$2.140) \\
                        
                        \textbf{Difference} & 0.45 ($\pm$1.154) & -0.61 ($\pm$2.681) & 0.32 ($\pm$1.308) & 5.47 ($\pm$8.845) & 0.46 ($\pm$1.477) \\
                        \bottomrule
                    \end{tabular}
                    \label{tab:tsa_sfts_vs_nettisa}
                \end{center}
            \end{table*}

        \subsubsection{Multiclass classification using merged datasets}
            %44 classes
            Since the novel universal Enhanced NeTtiSa features vector performed well in all evaluated cases---either binary or multiclass, we wanted to challenge the discriminative performance even more. Therefore, we merged the dataset to create a more complex problem with more classification classes. At first, we experiment with merging all datasets used for the binary classification problem. The resulting dataset contains ten classes\footnote{Botnet, Clear, Cryptomining, DNS Malware, DoH, DoS, IoT Malware, HTTPS Brute Force, Tor, and VPN}. The results of classification are presented in \tabref{tab:multiclass_all}. We achieve 97.14\% of the weighted average F1-score and 91.88\% of the macro average F1-score.

            \begin{table*}[t!]
                \caption{Final results (in \%) from the testing phase of multiclass classification based on NetTiSA flow of all binary problems together.}
                \begin{center}
                    \begin{tabular}{YXXXR}
                        \toprule 
                        \textbf{Class}
                         & \textbf{Precision} & \textbf{Recall} & \textbf{F1-score} & \textbf{Support}\\
                        \toprule
                        Botnet   & 99.36  &  96.66 &  97.99   & 12,751  \\
                        \rowcolor{LigthGray} Clear   & 97.55  &  96.82 &  97.18 & 5,283,309  \\
                        Cryptomining   & 99.73  &  99.52 &  99.62  & 368,528  \\
                        \rowcolor{LigthGray} DNS malware   & 90.79  &  76.67 &  83.13      & 90  \\
                        DoH   & 92.39  &  94.98 &  93.67 & 1,844,524  \\
                        \rowcolor{LigthGray} DoS   & 98.70  &  91.33 &  94.87   & 20,711  \\
                        IoT malware   & 99.80  &  99.23 &  99.51   & 13,287  \\
                        \rowcolor{LigthGray} HTTPS Brute Force   & 93.60  &  89.89 &  91.70  & 200,484  \\
                        Intrusion   & 98.03  &  98.06 &  98.05 & 2 247,509  \\
                        \rowcolor{LigthGray} TOR   & 85.80  &  65.07 &  74.01     & 418  \\
                        VPN   & 93.71  &  71.25 &  80.95    & 8,389  \\
                        \bottomrule
                        \textbf{Accuracy} & --   &  --  &  96.69 & 10,000,000 \\
                        \textbf{Macro avg.} &  95.41 &   89.04  &  91.88 & 10,000,000 \\
                        \textbf{Weighted avg.} &  96.71 &   96.69 &   96.69 & 10,000,000 \\
                        \bottomrule
                    \end{tabular}
                    \label{tab:multiclass_all}
                \end{center}
            \end{table*}

            Moreover, we merged all multiclass problems with several binary problems (detection of cryptomining, DNS malware, DoH, HTTPS Brute Force) into a single dataset. The created dataset contains 44 classification classes (see \ref{appendix:multiclass_multiclass}). The achieved results are shown in \tabref{tab:multiclass_all_multiclass}.  The classifier achieved 94.99\% of the weighted average F1-score and 71.81\% of the macro average F1-score.

            In both cases, the classifier performed well despite the large imbalance of the classes in the dataset---which explains the lower performance in the macro average F1-score. Nevertheless, the NetTiSA features achieve high accuracy even on fine-grained classification tasks with many classes.

            \begin{table*}
             \caption{Final results (in \%) from the testing phase of multiclass classification based on NetTiSA flow of all multiclass problems together. The crated dataset of all multiclass problems contains 44 classes.}
                \begin{center}
                    \begin{tabular}{YXXXR}
                        \toprule 
                         & \textbf{Precision} & \textbf{Recall} & \textbf{F1-score} & \textbf{Support}\\
                        \toprule
                        \textbf{Accuracy} & --   &  --  & 95.08  & 9,999,803 \\
                        \textbf{Macro avg.} & 83.49  & 66.51 & 71.81 & 9,999,803 \\
                        \textbf{Weighted avg.} & 95.03 & 95.08 & 94.99 & 9,999,803 \\
                        \bottomrule
                    \end{tabular}
                    \label{tab:multiclass_all_multiclass}
                \end{center}
            \end{table*}

    \subsection{Enhanced NetTiSA feature importances}
\begin{figure*}[t!]
                \centerline{\includegraphics[width=16cm]{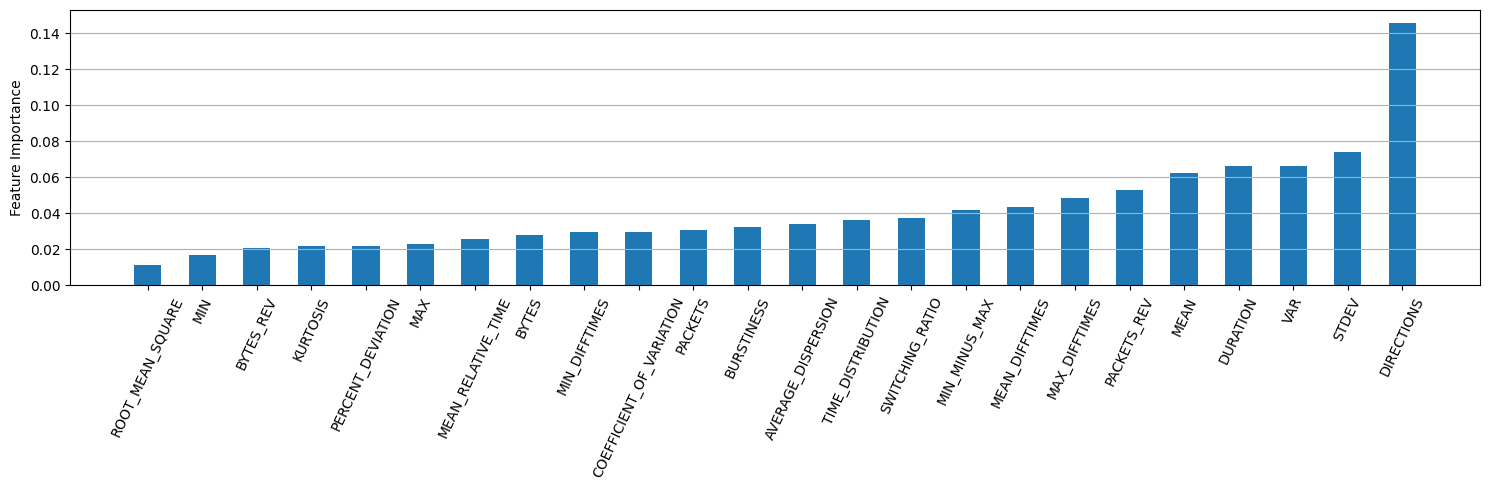}}
                \caption{Average feature importance across all classification problems measured by gain importance metric.}
                \label{fig:mean_feature_importances}
        \end{figure*}

        We extracted the feature importances of individual Enhanced NetTiSA features measured by the gain metric\footnote{the average gain across all splits the feature is used in} from each trained model. Nevertheless, the importance differs widely on each evaluated classification task; therefore we average the values across all cases. The average feature importances are presented in~\figref{fig:mean_feature_importances}. In the top ten most important features, there are mainly time-based features.

        Some classification tasks strongly influenced the feature importance of a single particular feature. For example, the high importance of the feature \textit{Directions} is caused by the Botnet, (D)DoS, IDS using the CIC dataset, and VPN (VNAT dataset) binary classification problems. Since in these problems is the asymmetry of the traffic extremely important for classification, the \textit{Directions} feature separates the majority of the dataset samples. 

        Other binary classification problems use primarily one or two features. Nevertheless, all multiclass problems use almost all features with small differences in their importance. The detailed view on feature importances for each classification task is shown in \ref{sec:feature_importance}.

    \subsection{Discussion about Feature Discriminative Performance}
        Despite the expected performance drops, due to the universality of the feature set, we consider the Discriminative Performance of Enhanced NetTiSA features as very good and comparable with the best-performing proposals from the state-of-the-art. And in only four cases out of 23, the classifier performed more than 1\% worse than the related work. Nevertheless, the reason behind the reduced accuracy was always investigated and explained. The higher accuracy in related works often stemmed from large telemetry sizes or higher complexity of feature computations.
        
\section{Network telemetry size comparison} \label{sec:telemetry_comparison}
        The size of the input feature vector matters, especially in commercial deployment. A larger feature vector puts additional stress on the monitoring infrastructure---the collectors and detectors must process more data. Moreover, in the case of flow monitoring infrastructure~\cite{Hofstede2014}, the large telemetry also utilizes bandwidth, that could be otherwise used by customers. Therefore, we compared the sizes of network telemetry (in our case, extended flows) used as the input for the classification across all the concerned tasks.

        To form a reference point, we need to establish the size of a classical IP flow record. We define a classical IP flow as a combination of flow key (Source IP, Destination IP, Source port, Destination port, Protocol) and a tuple of 6 values (Number of packets and bytes in the flow in both directions, first and last packet timestamp).  Thus, the transfer of this classical IP flow requires 21 bytes for the flow key and 40 bytes for the 6-value tuple---resulting in 61 bytes.

        Most of the approaches extend the flow for additional data, which naturally results in the size increase. We analyzed the flow records used by the best-performing classifiers on each evaluated classification task and established their input flow record size. When the datatypes were not specified, we assumed four-byte integers and four-byte floats for the record size calculation.

\begin{figure*}[t!]
                \centerline{\includegraphics[width=16cm]{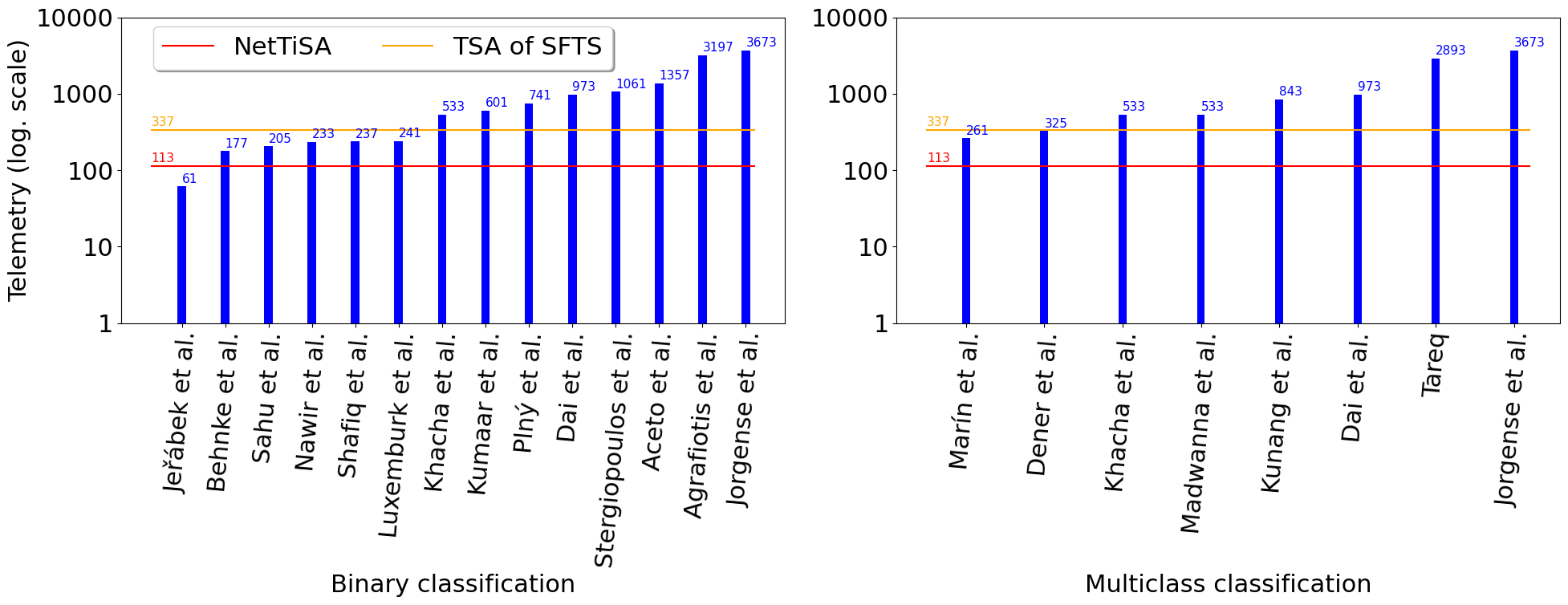}}
                \caption{Comparison of telemetry size of best-related works, complete feature set based on Time Series Analysis of Single Flow Time Series (TSA of SFTS) from our previous work \cite{koumar2023}, and NetTiSA flow.}
                \label{fig:telemetry_comparsion}
        \end{figure*}

        The results of our analysis are presented in~\figref{fig:telemetry_comparsion}. It can be seen, that NetTiSA requires almost the smallest flow records of 113 bytes (an increase of only 52 bytes compared to classical IP flow records). Most of the approaches require at least two times larger flow extension. The only approach that uses smaller flow records is the Jeřábek et al.~\cite{doh_kamil}, who compensate for the small amount of information with active probing. 
    
\subsection{Discussion about the telemetry size}
        The~\figref{fig:telemetry_comparsion} also shows, that most approaches do not consider a telemetry size as an important factor. For example, the Jorgense et al.~\cite{vnat_dataset} on the VNAT dataset requires 3,673 bytes on a single flow record---their records would consume around 29\,Gb of bandwidth per millions of flow. Such bandwidth requirement makes the approach hardly deployable since the large ISP network can generate more than millions of flow records per second. Compared to 29\,Gb, NetTiSA flow consumes only 904\,Mb per million flows. The traditional IP flow record requires only 488\,Mb per million flows. Nevertheless, since the NetTiSA flow bandwidth fits into 1\,Gbps lines, even with 1 million of flow per second, we consider its telemetry size as acceptable even in large infrastructure.

\section{The impact of NetTiSA features calculation} \label{sec:implementation}
    To establish the feasibility of NetTiSA features in high-speed infrastructures, we also measured the performance impact of NetTiSA flow features calculation. Therefore, we implemented the feature calculation into flow exporter ipfixprobe\footnote{https://github.com/CESNET/ipfixprobe}. The ipfixprobe is a high-performance flow exporter written in C++, that is capable of monitoring 100\,Gbps backbone lines. It supports multiple input types, from traditional slow interfaces such as libpcap to high-speed DPDK (Data Plane Development Kit) \cite{dpdk} inputs for ISP-based monitoring. 

    We leveraged the plugin-based architecture of the ipfixprobe and implemented NetTiSA flow as the process plugin with the same name. For each flow, the plugin performs the stream-wise computation of each feature described in~\secref{exported_features}. 

    Regarding operation memory, the NetTiSA feature computation stores 15 floats in the memory for each flow, resulting in 60 bytes per flow. Moreover, with the arrival of each new packet, the plugin performs 23 mathematical operations and five comparisons. Additionally, it performs 20 mathematical operations before exporting the flow.

    For the comparison, we measured the ipfixprobe exporter without the flow extensions (classical flows), and with three flow extensions: 1.\,the flow containing SPLT of length 30 packets (the \textit{pstats} plugin in ipfixprobe), 2.\,proposed NetTiSA flow, and 3.\,we also implemented as an additional ipfixprobe process plugin, that computes the time series analysis features presented by Koumar et al.~\cite{koumar2023}. The plugins have been tested in two environments. In the lab-traffic environment, we performed a stress test and tested the limits of the exporters using artificially created data. In the real-traffic environment, we utilized anonymized pcaps captured at the backbone ISP network and tested the exporters using real network data.

%    \begin{figure*}[t!]

%        \centering
%        \includegraphics[width=16cm]{pictures/deployment_of_nettisa_into_ISP_network.pdf}
%        \caption{The deployment of monitoring infrastructure with the \textit{NetTiSA flow} into ISP network. The network telemetry of each flow is the number of bytes sent from the Monitoring probe to the Collector through the network infrastructure.}
%        \label{fig:deployment}
%    \end{figure*}

%    Additionally, we also measured the operational memory requirements. First, we measure the memory requirements using \textit{Spirent test server} setup shown in \figref{fig:measuring-setup}. Second, we used three types of PCAP files for memory utilization: 1)\,Short flows -- only flows with 30 or fewer packets, i.e., flows that are fully captured by SPLT, 2)\,Long flows -- only flows with more than 30 packets, 3)\,Anonymized and payload-less CESNET2 captures -- captures of real office environment of subnetwork inside ISP CESNET2 network.

    \subsection{Performance with lab-traffic environment} \label{perf-contrains}
    
\begin{figure}
            \centering
            \includegraphics[width=0.65\textwidth]{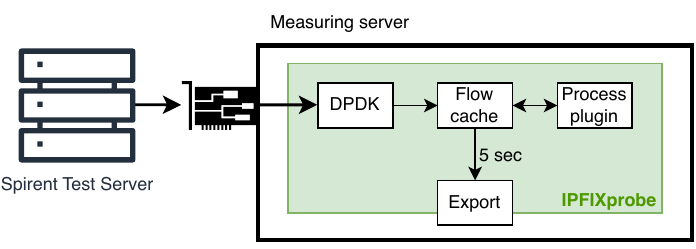}
            \caption{Measuring setup}
            \label{fig:measuring-setup}
        \end{figure}

        First, we measure the performance with the lab-traffic environment. For this purpose, we utilized the Spirent test server. The Spirent test server generates the network traffic according to the specified parameters and thus performs stress tests of network devices. The packets were transmitted to the server with ipfixprobe exporter, instantiated with the DPDK input plugin to allow high-speed packet processing. The measuring setup is shown in \figref{fig:measuring-setup}. 
        
        As mentioned above, we are only interested in the overall performance depending on flow length. Therefore, we defined three constraints to remove the influence of the input part of the exporter and flow cache as much as possible:
        \begin{enumerate}
            \item \textbf{Used uniformly randomized packet sizes} from 80 to 1400 bytes to maintain variable data for the processing.
            \item \textbf{Limited the traffic to only a single network flow.} The single flow stress test removes the influence of other parts in the exporter (such as cache misses, and collisions) and allows us to focus on the processing plugin performance and memory requirements. We force the flow cache to flush every five seconds to ensure that more than one flow is exported during the measurement. The measurement time is always 60 seconds; thus, the count of the exported flows should be exactly 12 ($60/5 = 12$). The number of packets/bytes in flow is then regulated by setting frames generated per second.
            \item \textbf{Set only one input DPDK thread} since we measured in the environment where a specific single flow is always mapped to the same DPDK queue. Thus, using multiple threats would be redundant.
        \end{enumerate}

%        We have defined several constraints to remove the influence of the input part of the exporter and flow cache as much as possible; we have: 1)\,used uniformly randomized packet size from 80 to 1400 bytes, 2)\,limited the traffic to only single network flow, and the export is ensured by a relatively short timeout (5 seconds) from the flow cache, since the flow length is important, instead of the number of flows, 3)\,limited the input to only one DPDK queue, i.e. one input thread.
        
%        The measuring setup is shown in \figref{fig:measuring-setup}. As mentioned above, we are only interested in the overall performance depending on flow length. We use variable packet lengths to prevent the potential effect of the packet size. Flow key values (IP addresses, protocol and ports) remain the same; thus all traffic is only one flow. The single flow stress test removes the influence of other parts in the exporter (such as cache misses, and collisions) and allows us to focus on the processing plugins performance and memory requirements. For that reason, it is unnecessary to use more DPDK queues, which would only cause the overhead (we measured in the environment where specific flow is always mapped to the same DPDK queue). To ensure we export more flows during the measurement, we essentially limited the validity of the flow cache entry to 5 seconds. The measurement time is always 60 seconds; thus, the count of the exported flows should be exactly 12 ($60/5 = 12$). The flow length is then regulated by setting frames generated per sec.

        \subsubsection{Computational performance}

            \begin{figure*}[t!]
                \centering
                \includegraphics[width=12cm]{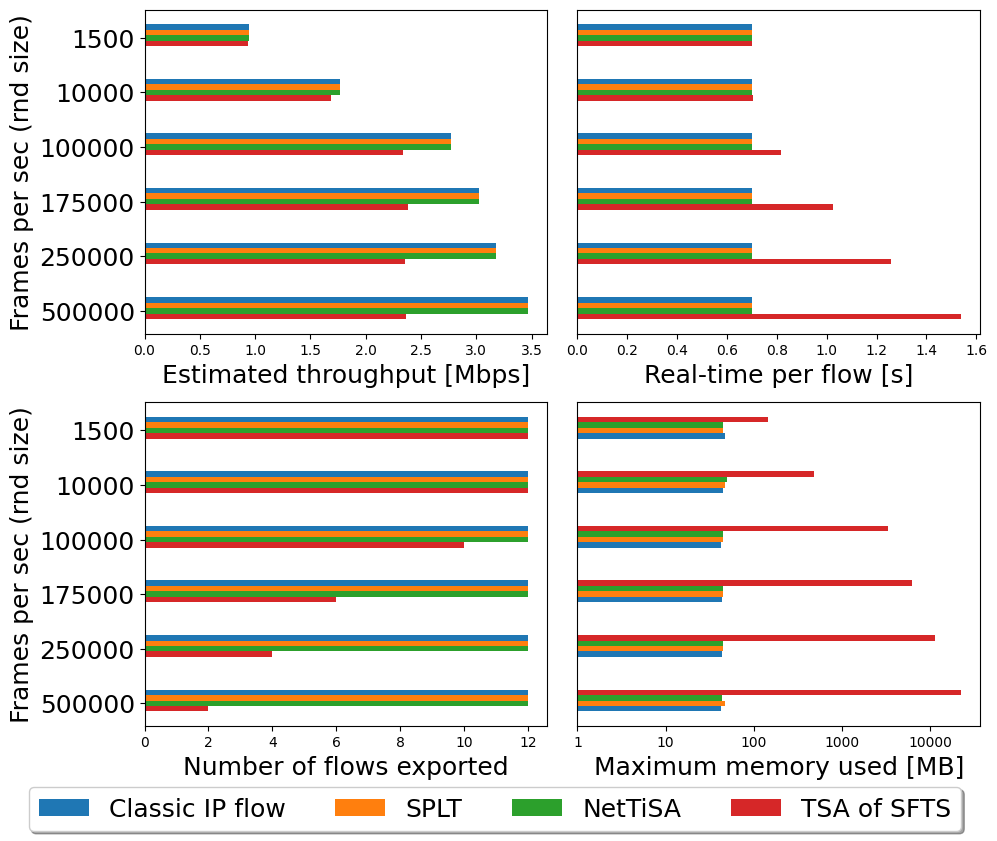}
                \caption{Results of performance and memory requirements comparison using Spirent test server}
                \label{fig:performace_spirent}
            \end{figure*}
            
            % \begin{figure*}[t!]
            %     \centering
            %     \includegraphics[width=\textwidth]{pictures/performance_spirent.png}
            %     \caption{Results of performance comparison using Spirent test server}
            %     \label{fig:performace_spirent}
            % \end{figure*}
    
            As we can see in \figref{fig:performace_spirent}, the exporting of NetTISA and SPLT features are comparable to classic IP Flow, and we do not observe any performance drop. On the other hand, the plugin TSA of SFTS lags behind (performance-wise) in every perspective. In the left top chart, we estimated the maximal throughput of the exporter based on the used plugin. The exporter is saturated relatively early in terms of frames per second---the input part of the exporter is saturated at approximately 500,000 frames per second (since we applied limitations described in \ref{perf-contrains}). After that, we cannot see any throughput gain. However, we can see that the TSA of SFTS causes the saturation much earlier and we cannot get a better throughput than around 200 Mbps. 
    
            In the right top chart, we used a metric called real-time per flow, which is the time of single flow processing. It cannot be less than 5 seconds because of flow-cache timeout. As we see, all plugins except TSA of SFTS behave reasonably---their processing time does not excessively exceed 5 seconds.
    
            The left bottom chart is a complement to the second chart. It shows how many flows the plugin can export---the maximum is 12, which means only the TSA of SFTS cannot export all available flows when frames per second are equal to or larger than 100,000.

        \subsubsection{Memory requirements}
        
            % \begin{figure*}[t!]
            %     \centering
            %     \includegraphics[width=0.5\textwidth]{pictures/ram_spirent.png}
            %     \caption{Maximum memory  }
            %     \label{fig:memory_spirent}
            % \end{figure*}

            In the right bottom chart of \figref{fig:performace_spirent} we see the memory performance follows the computational performance regarding trends across all measured plugins. This measurement method was based on recording the maximum memory allocated by the exporter. Thus, besides the actual measured plugin, the overhead of the exporter was included. Since the processed data was identical, this overhead can be assumed to be comparable and thus detectable in the whole measurement.

        \subsection{Performance on captured real traffic}
            For performance measurement on captured real traffic, we used three types of PCAP files for memory utilization: 1.\,Short flows -- only flows with 30 or fewer packets, i.e., flows that are fully captured by SPLT, 2.\,Long flows -- only flows with more than 30 packets, 3.\,Anonymized and payload-less CESNET2 captures -- captures of real office environment of subnetwork inside ISP CESNET2 network.
        
        \subsubsection{Computational performance}
            The performance comparison in processing time for each feature type computation is shown in \tabref{tab:performance_comparison_by_pcap}. We can see expected behavior in performance on the real traffic captures. It must be noted that the processing time is influenced by loading packets from the PCAP from the file (the processing time of the same number of packets in deployment into a real network environment is much faster). The Classical IP flow has the best performance because it performs the least operations, the SPLT flow performs extra assignments into memory (max 30 times), and the NetTiSA flow performs several extra mathematical operations when a new packet arrives. Thus, these three types of flows are sorted by performance as: 1.~Classical IP, 2.~SPLT, and 3.~NetTiSA flow. However, the difference in performance between them is not significant. Furthermore, the performance of the computation of features generated by the TSA of SFTS needs a lot of computation, resulting in a significant decrease in performance compared to other types of flows.
    
            \begin{table}[h]
                \caption{Performance comparison on real traffic by using captured PCAPs.}
                 \begin{center}
                    % \begin{tabular}{l c c c c c}
                    %     \toprule
                    %     &  & \multicolumn{4}{c}{\textbf{Processing time}} \\
                    %     \cmidrule{3-6}
                    %    \textbf{PCAP description} & \textbf{Size of PCAP} & \textbf{Classic IP flow} & \textbf{SPLT flow} & \textbf{NetTiSA flow} & \textbf{TSA of SFTS} \\
                    %     \toprule
                    %     Short flows     & 299M  &  1.1    &  1.0    &  1.2    & 67.0 \\
                    %     Long flows      & 13G   & 46.4   & 45.4   & 42.5   & 877.1 \\
                    %     CESNET2 capture & 2G    &  7.7    &  8.1    &  9.6    & 523.1 \\
                    %     \bottomrule
                    \begin{tabular}{l | c c | c c c c}
                        \toprule
                        &  \multicolumn{2}{c|}{\textbf{Number of}} & \multicolumn{4}{c}{\textbf{Processing time [s] of}} \\
                        % \cmidrule{4-7}
                       \multirow{-2}{*}{\textbf{PCAP}}  & \textbf{packets} & \textbf{flows} & \textbf{Classic IP} & \textbf{SPLT} & \textbf{NetTiSA} & \textbf{TSA of SFTS} \\
                        \toprule
                        Short flows      & 3.3m  & 24,6k  & 1.0    &  1.0    &  1.2    & 67.0 \\
                        Long flows       & 139.2m  & 1.3k   & 40.4   & 44.3   & 45.0  & 877.1 \\
                          & 26.6m & 213.3k & 7.7    &  8.1    &  9.6    & 523.1 \\
                        \multirow{-2}{*}{CESNET2 captures}  & 130m & 1m & 38.5 & 41.7 & 47.6 & 2563.9 \\
                        \bottomrule
                    \end{tabular}
                    \label{tab:performance_comparison_by_pcap}
                \end{center}
            \end{table}

        % \begin{table}[h]
        %         \caption{}
        %          \begin{center}
        %             \begin{tabular}{l r r r}
        %                 \toprule
        %                 & \multicolumn{3}{c}{\textbf{PCAP}} \\
        %                 \cmidrule{2-4}
        %                 & Short flows & Long flows & CESNET2 capture \\
        %                 \cmidrule{2-4}
        %                 Size of PCAP & 299M & 13G & 2G \\
        %                \toprule
        %                \textbf{Flow} & \multicolumn{3}{c}{\textbf{Processing time}} \\
        %                 \toprule
        %                 \textbf{Classic IP}    & 1.1041     & 46.4960   & 7.7287 \\
        %                 \textbf{SPLT}        & 1.0037     & 45.4922   & 8.1291 \\
        %                 \textbf{NetTiSA}       & 1.2045     & 42.5661   & 9.6354 \\
        %                 \textbf{TSA of SFTS}   & 67.0785    & 877.1240  & 523.1186 \\
        %                 \bottomrule
        %             \end{tabular}
        %             \label{tab:performance_comparison_by_pcap_2}
        %         \end{center}
        %     \end{table}

    \subsubsection{Memory requirements}

        The results of the memory requirement of each set of feature computations are shown in \tabref{tab:memory_comparison_by_pcap}. Similarly to performance measurement, we also used packet captures (pcaps) from a real CESNET2 ISP network. The results are provided by mean and standard deviation of memory usage per second. We can see several points from the results: 1.\,The number of packets in flow does not influence the memory requirements of the feature computation of Classical IP, SPLT, and NetTiSA flows. 2.\,The number of flows does not highly influence the memory requirements of the feature computation of Classical IP and NetTiSA flows. However, the SPLT flow is influenced by it. 3.\,The TSA of SFTS feature computation has memory requirements that are highly influenced by the number of packets in the flow (saving the time series into memory requires dynamic allocation). 4.\,The standard deviation shows that Classical IP, SPLT, and NetTiSA flows are highly stable in memory requirements, but the TSA of SFTS is highly unstable because of free large vectors of payload length and times after exportation of the flow. 5.\,The difference between the memory requirements of the TSA of SFTS on Short and Long flows is huge (from 26.5 MiB on 1k of flows to 5.24 GiB on 1k of flows).

        % 1. závislost mezi počtem paketů a používáním ram basic, SPLT, nettisa není (z long flows to dje vidět kde je mrte paketů)
        
        % 2. závislost meti počtem flow a používáním ram není u basic a nettisa, ale je u SPLT

        % 3. tsa of sfts silně závisí na počtu paketů

        % 4. z smerodatne odchylky je vidět stabilita používání ram v čase u basic, SPLT a nettisa, naopak tsa of sfts je silne nestabilní (způsobeno, jak se uvolňují 2 velké vektory po exportování jednoho flow)

        % 5. Rozdíl mezi Long flows a Short flows se nejvíce projevuje u TSA of SFTS, kde použitá memory na jeden flow exponencionálně vzroste (z 26.5 MiB na 1k flow na 5.24 GiB na 1k flow) 

        \begin{table}[h]
                \caption{The comparison of memory usage per second of input PCAPs processing. The comparison is done by mean and standard deviation.}
                 \begin{center}
                    \begin{tabular}{l | c c | c c c c}
                        \toprule
                        &  \multicolumn{2}{c|}{\textbf{Number of}} & \multicolumn{4}{c}{\textbf{Mean memory usage per second [MiB] of}} \\
                        % \cmidrule{4-7}
                       \multirow{-2}{*}{\textbf{PCAP}} & \textbf{packets} & \textbf{flows} & \textbf{Classic IP} & \textbf{SPLT} & \textbf{NetTiSA} & \textbf{TSA of SFTS} \\
                        \toprule
                        Short flows     & 3.3m  & 24,6k  & 30.2 ($\pm$0.0) & 36.8 ($\pm$3.1) & 31.0 ($\pm$0.4) & 653 ($\pm$615) \\
                        Long flows      & 139.2m  & 1.3k   & 30.1 ($\pm$0.0) & 30.5 ($\pm$0.3) & 30.2 ($\pm$0.0) & 6982 ($\pm$7093) \\
                         & 26.6m & 213.3k & 30.1 ($\pm$0.0) & 40.9 ($\pm$1.6) & 31.5 ($\pm$0.2) & 2632 ($\pm$4758) \\
                        \multirow{-2}{*}{CESNET2 captures}  & 130m & 1m & 30.2 ($\pm$0.0) & 41.7 ($\pm$0.8) & 31.6 ($\pm$0.1) & 3847 ($\pm$5855) \\
                        \bottomrule
                    \end{tabular}
                    \label{tab:memory_comparison_by_pcap}
                \end{center}
            \end{table}

    \subsection{Discussion about the performance}
        Since the NetTiSA feature computation performance was almost identical to the traditional flow computation, we consider its performance excellent. Moreover, the NetTiSA plugin has already been deployed in the production CESNET2 network with eight monitoring probes, with approximately 200 thousand flows each second. Thus, we assume the NetTiSA is deployment-ready for other flow-based monitoring infrastructure.

\section{Conclusion} \label{conclusion_section}
    In this paper, we proposed a novel extended IP flow called \textit{NetTiSA} flow that is built on Time Series Analysis of Single Flow Time Series. The \textit{NetTiSA} flow contains 13 features based on statistics of payload lengths, statistics of transmission times, and distribution and behavior of packets within a flow. Moreover, an additional seven features can be computed from 13 exported features, resulting in 20 features called \textit{Enhanced NetTiSA} flow. The purpose of additional features is to improve the classification performance.  The usability of \textit{Enhanced NetTiSA} flow was thoroughly evaluated from three main perspectives: 1.\,Discriminative performance, 2.\,Flow telemetry size and bandwidth requirements, and 3.\,Feature computational overhead.
    
    The discriminative performance of the features carried by \textit{Enhanced NetTiSA} flow was evaluated on 25 network classification tasks using 15 publicly available and well-known network traffic datasets which are often used in recent research. These datasets were then used to train and evaluate ML models and compare their performance to the best-performing classifiers from related works. In our design and evaluation pipeline, we created 2,500 models across multiple binary and multiclass classification tasks and showed the universality of the proposed features. 

    Furthermore, we prepared the C++ implementation of the NetTiSA feature calculation inside flow exporter \textit{ipfixprobe}. Moreover, the implementation of the feature extraction is developed in the form of a library, which allows fast integration into other flow-monitoring software.
    The implementation of NetTiSA flow feature extraction was thoroughly evaluated for speed and memory utilization.  Our experiments showed that the \textit{NetTiSA} flow has negligible impact on the performance of the flow monitoring probe; thus it can process 100\,Gbps of network traffic.
    
    The best-performing related works were analyzed for their flow-record sizes to discuss their usability in real-world deployment. Since the NetTiSA flow records can be exported via a single 1\,Gbps line on a network with 1 million flows per second, we considered the size of flow records as sufficient for deployment even in large ISP networks. The NetTiSA flow deployment does not significantly increase the required bandwidth for flow monitoring while enabling accurate traffic classification.

    Overall, the novel Enhanced NetTiSA flow feature vector has the potential to replace the industry standard SPLT and other flow extension approaches, due to its lightweight extraction, small flow size, and excellent discrimination capability across multiple network traffic classification tasks. Moreover, in many cases, the trained classifiers outperformed the current state-of-the-art. This shows the Enhanced NetTiSA flow feature vector usability for network traffic classification in large ISP networks.

\section*{Acknowledgment}
    This research was funded by the Ministry of Interior of the Czech Republic, grant No. VJ02010024: Flow-Based Encrypted Traffic Analysis and also by the Grant Agency of the CTU in Prague, grant No. SGS23/207/OHK3/3T/18 funded by the MEYS of the Czech Republic.

\appendix

\section{Complete results of classification based on Enhanced NetTiSA flow} \label{sec:complere_results}

 \begin{table}[H]
        \caption{Comparison of results of multiclass classification based on Enhanced NetTiSA flows with best-related work on the same dataset. We compare results by Accuracy in \%, Macro average F1-score in \%, and Weighted average F1-score in \%. The Telemetry column represents the size of the flow extension (the classical flow has an extension size equal to zero). The green background color marks fields where classification based on the NetTiSA flow is at least 1\% better than best-related work. The red background color marks fields where classification based on the NetTiSA flow is at least 1\% worse than best-related work. The grey background color marks the rest fields.}
        \begin{center}
            \begin{tabular}{|l|l|l|r|c|c|c|}
                \hline
                \rowcolor{Gray} &  &  &  &  & \textbf{Macro} & \textbf{Weighted} \\
                \rowcolor{Gray} \multirow{-2}{*}{\textbf{Task}} & \multirow{-2}{*}{\textbf{Dataset}} &  \multirow{-2}{*}{\textbf{Approach}} & \multirow{-2}{*}{\textbf{Telelemetry}} & \multirow{-2}{*}{\textbf{Accuracy}} & \textbf{avg. F1} & \textbf{avg. F1} \\
                \hline
                 % &  &  &  &  \\ 
                & & \cellcolor[HTML]{BBFFBB}Marín et al. \cite{8844609} & \cellcolor[HTML]{BBFFBB}200 & \cellcolor{LigthGray}99.72 & \cellcolor[HTML]{BBFFBB}76.04 &  \cellcolor[HTML]{BBFFBB}98.00 \\
                \multirow{-2}{*}{Botnet} &  &  \cellcolor[HTML]{BBFFBB}TSA of SFTS   \cite{koumar2023} & \cellcolor[HTML]{BBFFBB}276 & \cellcolor{LigthGray}99.73 & \cellcolor[HTML]{BBFFBB}82.79 & \cellcolor[HTML]{BBFFBB}99.73 \\
                \multirow{-2}{*}{class.} & \multirow{-3}{*}{CTU-13 \cite{GARCIA2014100}} &  \cellcolor[HTML]{BBFFBB}NetTiSA  & \cellcolor[HTML]{BBFFBB}\textbf{52} & \cellcolor{LigthGray}99.85 & \cellcolor[HTML]{BBFFBB}\textbf{89.31} & \cellcolor[HTML]{BBFFBB}\textbf{99.84} \\
                 
                \hline
                % \cline{2-7}
                
                & & \cellcolor[HTML]{BBFFBB}Khacha et al. \cite{khacha2022hybrid}  & \cellcolor[HTML]{BBFFBB}472 & \cellcolor[HTML]{BBFFBB}98.69 & \cellcolor[HTML]{BBFFBB}-- &  \cellcolor[HTML]{BBFFBB}-- \\
                &  &  \cellcolor[HTML]{BBFFBB}TSA of SFTS   \cite{koumar2023} & \cellcolor[HTML]{BBFFBB}276 & \cellcolor[HTML]{BBFFBB}\textbf{99.97} & \cellcolor[HTML]{BBFFBB}\textbf{89.75}  & \cellcolor[HTML]{BBFFBB}\textbf{99.97}\\
                & \multirow{-3}{*}{Edge-IIoTset \cite{mbc1-1h68-22}} &  \cellcolor[HTML]{BBFFBB}NetTiSA  & \cellcolor[HTML]{BBFFBB}\textbf{52} & \cellcolor[HTML]{BBFFBB}\textbf{99.96} & \cellcolor[HTML]{BBFFBB}\textbf{89.01} & \cellcolor[HTML]{BBFFBB}\textbf{99.96} \\

                % \hline
                \cline{2-7}
                
                \multirow{-4}{*}{IoT} & & \cellcolor{LigthGray}Tareq et al.  \cite{tareq2022analysis}   & \cellcolor[HTML]{BBFFBB}2,832 &  \cellcolor[HTML]{FAE5E4}\textbf{98.50} & \cellcolor[HTML]{BBFFBB}52.20 & \cellcolor[HTML]{FAE5E4}\textbf{98.57} \\
                \multirow{-4}{*}{Malware} & &  \cellcolor{LigthGray}TSA of SFTS   \cite{koumar2023} & \cellcolor[HTML]{BBFFBB}276 & \cellcolor[HTML]{FAE5E4}97.53 & \cellcolor[HTML]{BBFFBB}81.02  & \cellcolor[HTML]{FAE5E4}97.51 \\
               \multirow{-4}{*}{class.} & \multirow{-3}{*}{TON\_ IoT \cite{moustafa2021new}} &  \cellcolor{LigthGray}NetTiSA  & \cellcolor[HTML]{BBFFBB}\textbf{52} & \cellcolor[HTML]{FAE5E4}96.06 & \cellcolor[HTML]{BBFFBB}\textbf{82.82} & \cellcolor[HTML]{FAE5E4}96.01 \\

                \hline
                 
               & &  \cellcolor[HTML]{BBFFBB}Kunang et al. \cite{kunang2021attack}   & \cellcolor[HTML]{BBFFBB}784 &  \cellcolor[HTML]{BBFFBB}95.79 & \cellcolor[HTML]{BBFFBB}84.54 &  \cellcolor[HTML]{BBFFBB}95.11 \\
                & &  \cellcolor[HTML]{BBFFBB}TSA of SFTS   \cite{koumar2023} & \cellcolor[HTML]{BBFFBB}276 & \cellcolor[HTML]{BBFFBB}\textbf{99.93} & \cellcolor[HTML]{BBFFBB}83.23  & \cellcolor[HTML]{BBFFBB}\textbf{99.92} \\
               & \multirow{-3}{*}{CIC-IDS-2017 \cite{sharafaldin2018toward}} &  \cellcolor[HTML]{BBFFBB}NetTiSA  & \cellcolor[HTML]{BBFFBB}\textbf{52} & \cellcolor[HTML]{BBFFBB}\textbf{99.86} & \cellcolor[HTML]{BBFFBB}\textbf{88.70}  & \cellcolor[HTML]{BBFFBB}\textbf{99.85} \\

                 % \hline
                \cline{2-7}

               & &   \cellcolor[HTML]{BBFFBB}Madwanna et al. \cite{madwanna2023yars} & \cellcolor[HTML]{BBFFBB}472 & \cellcolor[HTML]{BBFFBB}82.21 &  \cellcolor[HTML]{BBFFBB}53.15 & \cellcolor[HTML]{BBFFBB}80.30 \\
               \multirow{-5}{*}{Intrusion} & &  \cellcolor[HTML]{BBFFBB}TSA of SFTS   \cite{koumar2023} & \cellcolor[HTML]{BBFFBB}276 & \cellcolor[HTML]{BBFFBB}95.60 & \cellcolor[HTML]{BBFFBB}40.22  & \cellcolor[HTML]{BBFFBB}95.08 \\
               \multirow{-5}{*}{class.} & \multirow{-3}{*}{UNSW-NB15 \cite{moustafa2015unsw}} &  \cellcolor[HTML]{BBFFBB}NetTiSA  & \cellcolor[HTML]{BBFFBB}\textbf{52} & \cellcolor[HTML]{BBFFBB}\textbf{98.85} & \cellcolor[HTML]{BBFFBB}\textbf{67.90} & \cellcolor[HTML]{BBFFBB}\textbf{98.78} \\
      
                \hline
                %    &  &  &  & --
                & &  \cellcolor{LigthGray}Dai et al.  \cite{dai2023glads} & \cellcolor[HTML]{BBFFBB}912 & \cellcolor[HTML]{FAE5E4}\textbf{97.95} & \cellcolor[HTML]{FAE5E4}\textbf{86.77} & \cellcolor{LigthGray}-- \\
               \multirow{-2}{*}{TOR} &  &  \cellcolor{LigthGray}TSA of SFTS   \cite{koumar2023} & \cellcolor[HTML]{BBFFBB}276 & \cellcolor[HTML]{FAE5E4}95.48 & \cellcolor[HTML]{FAE5E4}79.87  & \cellcolor{LigthGray}95.20 \\
                \multirow{-2}{*}{class.} & \multirow{-3}{*}{ISCX-Tor-2016 \cite{lashkari2017characterization}} &  \cellcolor{LigthGray}NetTiSA  & \cellcolor[HTML]{BBFFBB}\textbf{52} & \cellcolor[HTML]{FAE5E4}96.36 & \cellcolor[HTML]{FAE5E4}84.46 & \cellcolor{LigthGray}96.16 \\

                \hline

               & &   \cellcolor[HTML]{BBFFBB}Dener et al.   \cite{dener2023rfse}  & \cellcolor[HTML]{BBFFBB}264 & \cellcolor[HTML]{BBFFBB}89.29 & \cellcolor[HTML]{BBFFBB}87.83 & \cellcolor[HTML]{BBFFBB}90.49  \\
               &  &  \cellcolor[HTML]{BBFFBB}TSA of SFTS   \cite{koumar2023} & \cellcolor[HTML]{BBFFBB}276 & \cellcolor[HTML]{BBFFBB}\textbf{94.80} & \cellcolor[HTML]{BBFFBB}\textbf{91.21}  & \cellcolor[HTML]{BBFFBB}\textbf{94.77} \\
               & \multirow{-3}{*}{ISCX-VPN-2016 \cite{icxs_vpn_2016_dataset}} &  \cellcolor[HTML]{BBFFBB}NetTiSA  & \cellcolor[HTML]{BBFFBB}\textbf{52} & \cellcolor[HTML]{BBFFBB}\textbf{94.04} & \cellcolor[HTML]{BBFFBB}89.66 & \cellcolor[HTML]{BBFFBB}\textbf{94.00} \\     
                
                % \hline
                \cline{2-7}
                
               & &  \cellcolor[HTML]{BBFFBB}Jorgense et al.  \cite{vnat_dataset}  & \cellcolor[HTML]{BBFFBB}3,612 & \cellcolor[HTML]{BBFFBB}96 & \cellcolor[HTML]{BBFFBB}--  & \cellcolor[HTML]{BBFFBB}--  \\
              \multirow{-5}{*}{VPN} &  &  \cellcolor[HTML]{BBFFBB}TSA of SFTS   \cite{koumar2023} & \cellcolor[HTML]{BBFFBB}276 & \cellcolor[HTML]{BBFFBB}\textbf{98.60} & \cellcolor[HTML]{BBFFBB}\textbf{98.88}  & \cellcolor[HTML]{BBFFBB}\textbf{98.60} \\
               \multirow{-5}{*}{class.} & \multirow{-3}{*}{VNAT \cite{vnat_dataset}} &  \cellcolor[HTML]{BBFFBB}NetTiSA  & \cellcolor[HTML]{BBFFBB}\textbf{52} & \cellcolor[HTML]{BBFFBB}\textbf{99.29}  & \cellcolor[HTML]{BBFFBB}\textbf{98.92} & \cellcolor[HTML]{BBFFBB}\textbf{99.29} \\

                \hline
            \end{tabular}
            \label{tab:classification_results_mutliclass}
        \end{center}
    \end{table}

    \begin{table}[H]
        \caption{Comparison of results of binary classification based on Enhanced NetTiSA flows with best-related work on the same dataset. The Telemetry column represents the size of the flow extension (the classical flow has an extension size equal to zero). The green background color marks fields where classification based on the NetTiSA flow is at least 1\% better than best-related work. The red background color marks fields where classification based on the NetTiSA flow is at least 1\% worse than best-related work. The grey background color marks the rest fields.}
        \begin{center}
            \begin{tabular}{|l|l|l|r|c|c|}
                \hline
                \rowcolor{Gray} \textbf{Task} & \textbf{Dataset} &  \textbf{Approach} & \textbf{Telelemetry} & \textbf{Accuracy} & \textbf{F1-score} \\
                \hline
                &  &  \cellcolor{LigthGray}Stergiopoulos  et  al.   \cite{stergiopoulos2018automatic} & \cellcolor[HTML]{BBFFBB}1,000 & \cellcolor{LigthGray}99.85 & \cellcolor{LigthGray}99.90 \\
                \multirow{-2}{*}{Botnet} &  &  \cellcolor{LigthGray}TSA of SFTS   \cite{koumar2023} & \cellcolor[HTML]{BBFFBB}276 & \cellcolor{LigthGray}99.98 & \cellcolor{LigthGray}99.93 \\
                \multirow{-2}{*}{detection} &  \multirow{-3}{*}{CTU-13 \cite{GARCIA2014100}} & \cellcolor{LigthGray}NetTiSA   & \cellcolor[HTML]{BBFFBB}\textbf{52} & \cellcolor{LigthGray}99.95 & \cellcolor{LigthGray}99.79 \\
                
                \hline
                
                 &  & \cellcolor[HTML]{BBFFBB}Plný et al. \cite{plny2023decrypto} & \cellcolor[HTML]{BBFFBB}680 & \cellcolor[HTML]{BBFFBB}93.72 & \cellcolor[HTML]{BBFFBB}90.59 \\
                \multirow{-2}{*}{Cryptomining} &   &  \cellcolor[HTML]{BBFFBB}TSA of SFTS   \cite{koumar2023} & \cellcolor[HTML]{BBFFBB}276 & \cellcolor[HTML]{BBFFBB}95.29 & \cellcolor[HTML]{BBFFBB}93.11 \\
                \multirow{-2}{*}{detection} &  \multirow{-3}{*}{CESNET-MINER22 \cite{richard_plny_2022_7189293}} & \cellcolor[HTML]{BBFFBB}NetTiSA   & \cellcolor[HTML]{BBFFBB}\textbf{52} & \cellcolor[HTML]{BBFFBB}\textbf{97.32} & \cellcolor[HTML]{BBFFBB}\textbf{96.19}   \\
                
                \hline
                
                DNS &  & \cellcolor{LigthGray}Kumaar et al. \cite{kumaar2021hybrid}  & \cellcolor[HTML]{BBFFBB}540 & \cellcolor{LigthGray}99.19 & \cellcolor{LigthGray}99.20  \\
                Malware &   &  \cellcolor{LigthGray}TSA of SFTS   \cite{koumar2023} & \cellcolor[HTML]{BBFFBB}276 & \cellcolor{LigthGray}100.0 & \cellcolor{LigthGray}100.0 \\
                detection &  \multirow{-3}{*}{CIC-Bell-DNS \cite{cic_bell_dns_2021_article}} & \cellcolor{LigthGray}NetTiSA   & \cellcolor[HTML]{BBFFBB}\textbf{52} & \cellcolor{LigthGray}100.0 & \cellcolor{LigthGray}100.0   \\
                
                \hline
                
                &   &   \cellcolor{LigthGray}Behnke et al. \cite{behnke2021feature} & \cellcolor[HTML]{BBFFBB}116 & \cellcolor{LigthGray}-- & \cellcolor{LigthGray}99.8 \\
                &    &  \cellcolor{LigthGray}TSA of SFTS   \cite{koumar2023} & \cellcolor[HTML]{BBFFBB}276 & \cellcolor{LigthGray}99.90 & \cellcolor{LigthGray}99.84 \\
                &  \multirow{-3}{*}{CIC-DoH-Brw \cite{montazerishatoori2020detection}} & \cellcolor{LigthGray}NetTiSA   & \cellcolor[HTML]{BBFFBB}\textbf{52} & \cellcolor{LigthGray}99.89 & \cellcolor{LigthGray}99.83 \\
        
                % \hline
                \cline{2-6}
                
                 &  & \cellcolor{LigthGray}Jeřábek et al.~\cite{doh_kamil} & \cellcolor[HTML]{FAE5E4}\textbf{0} &  \cellcolor{LigthGray}97.5 &  \cellcolor{LigthGray}98.7  \\
                 \multirow{-5}{*}{DoH} &   &  \cellcolor{LigthGray}TSA of SFTS   \cite{koumar2023} & \cellcolor[HTML]{FAE5E4}276 & \cellcolor{LigthGray}97.79 & \cellcolor{LigthGray}98.80 \\
                \multirow{-5}{*}{detection} &  \multirow{-3}{*}{DoH-Real-world \cite{Jerabek2022}}  & \cellcolor{LigthGray}NetTiSA   & \cellcolor[HTML]{FAE5E4}52 & \cellcolor{LigthGray}97.79 & \cellcolor{LigthGray}98.80  \\
        
                \hline
                 
                 &   & \cellcolor{LigthGray}Luxemburk et al. \cite{9375998}  & \cellcolor[HTML]{BBFFBB}180 & \cellcolor{LigthGray}99.93 & \cellcolor[HTML]{BBFFBB}96.26  \\
                 \multirow{-2}{*}{Brute force} &    &  \cellcolor{LigthGray}TSA of SFTS   \cite{koumar2023} & \cellcolor[HTML]{BBFFBB}276 & \cellcolor{LigthGray}99.99 & \cellcolor[HTML]{BBFFBB}\textbf{99.83} \\
                 \multirow{-2}{*}{detection} &  \multirow{-3}{*}{HTTPS Brute-force \cite{jan_luxemburk_2020_4275775}} & \cellcolor{LigthGray}NetTiSA   & \cellcolor[HTML]{BBFFBB}\textbf{52} & \cellcolor{LigthGray}99.98 & \cellcolor[HTML]{BBFFBB}\textbf{99.80} \\
        
                \hline
                
                 &  &  \cellcolor{LigthGray}Shafiq  et  al. \cite{SHAFIQ2020433} & \cellcolor[HTML]{BBFFBB}176 & \cellcolor{LigthGray}99.99 & \cellcolor{LigthGray}99.99    \\
                 \multirow{-2}{*}{DoS} &  &  \cellcolor{LigthGray}TSA of SFTS   \cite{koumar2023} & \cellcolor[HTML]{BBFFBB}276 & \cellcolor{LigthGray}100.0 & \cellcolor{LigthGray}100.0 \\
                 \multirow{-2}{*}{detection} &  \multirow{-3}{*}{Bot-IoT \cite{dos_iot_Dataset}} & \cellcolor{LigthGray}NetTiSA   & \cellcolor[HTML]{BBFFBB}\textbf{52} & \cellcolor{LigthGray}100.0 & \cellcolor{LigthGray}99.98 \\
        
                \hline
                
                 &  &  \cellcolor[HTML]{BBFFBB}Sahu et al. \cite{sahu2021internet}  & \cellcolor[HTML]{BBFFBB}144 & \cellcolor[HTML]{BBFFBB}96 &  \cellcolor[HTML]{BBFFBB}96   \\
                 &   &  \cellcolor[HTML]{BBFFBB}TSA of SFTS   \cite{koumar2023} & \cellcolor[HTML]{BBFFBB}276 & \cellcolor[HTML]{BBFFBB}\textbf{99.86} & \cellcolor[HTML]{BBFFBB}\textbf{99.91} \\
                &  \multirow{-3}{*}{IoT-23 \cite{sebastian_garcia_2020_4743746}} & \cellcolor[HTML]{BBFFBB}NetTiSA   & \cellcolor[HTML]{BBFFBB}\textbf{52} & \cellcolor[HTML]{BBFFBB}\textbf{99.85} & \cellcolor[HTML]{BBFFBB}\textbf{99.90}  \\
        
                % \hline
                \cline{2-6}
                
                 &   &  \cellcolor{LigthGray}Khacha et al. \cite{khacha2022hybrid}  & \cellcolor[HTML]{BBFFBB}472  & \cellcolor{LigthGray}99.99 &  \cellcolor{LigthGray}99.99  \\
                 &    &  \cellcolor{LigthGray}TSA of SFTS   \cite{koumar2023} & \cellcolor[HTML]{BBFFBB}276 & \cellcolor{LigthGray}99.99 & \cellcolor{LigthGray}99.97 \\
                &  \multirow{-3}{*}{Edge-IIoTset \cite{mbc1-1h68-22}} & \cellcolor{LigthGray}NetTiSA   & \cellcolor[HTML]{BBFFBB}\textbf{52} & \cellcolor{LigthGray}100.0 & \cellcolor{LigthGray}99.98\\
        
                % \hline
                \cline{2-6}
                
                 \multirow{-7}{*}{IoT} &   &  \cellcolor{LigthGray}Dai et al.  \cite{dai2023glads}  & \cellcolor[HTML]{BBFFBB}912 & \cellcolor{LigthGray}99.29  & \cellcolor{LigthGray}99.03  \\
                 \multirow{-7}{*}{Malware} &    &  \cellcolor{LigthGray}TSA of SFTS   \cite{koumar2023} & \cellcolor[HTML]{BBFFBB}276 & \cellcolor{LigthGray}99.96 & \cellcolor{LigthGray}99.98 \\
                 \multirow{-7}{*}{detection} &  \multirow{-3}{*}{TON\_IoT \cite{moustafa2021new}} & \cellcolor{LigthGray}NetTiSA   & \cellcolor[HTML]{BBFFBB}\textbf{52} &  \cellcolor{LigthGray}99.95 & \cellcolor{LigthGray}99.97 \\
        
                \hline
                
                &  &  \cellcolor[HTML]{BBFFBB}Agrafiotis et al. \cite{Agrafiotis_Makri_Flionis_Lalas_Votis_Tzovaras_2022}  & \cellcolor[HTML]{BBFFBB}3,136 & \cellcolor[HTML]{BBFFBB}98.5 & \cellcolor[HTML]{BBFFBB}95.4 \\
                &   &  \cellcolor[HTML]{BBFFBB}TSA of SFTS   \cite{koumar2023} & \cellcolor[HTML]{BBFFBB}276 & \cellcolor[HTML]{BBFFBB}\textbf{99.89} & \cellcolor[HTML]{BBFFBB}\textbf{99.75} \\
               &   \multirow{-3}{*}{CIC-IDS-2017 \cite{sharafaldin2018toward}} & \cellcolor[HTML]{BBFFBB}NetTiSA   & \cellcolor[HTML]{BBFFBB}\textbf{52} &  \cellcolor[HTML]{BBFFBB}\textbf{99.75} & \cellcolor[HTML]{BBFFBB}\textbf{99.43} \\
        
                % \hline
                \cline{2-6}
                
                &   &   \cellcolor{LigthGray}Nawir et al. \cite{nawir2018performances}  & \cellcolor[HTML]{BBFFBB}172 & \cellcolor[HTML]{BBFFBB}94.37 &  \cellcolor[HTML]{FAE5E4}94.54  \\
                \multirow{-5}{*}{Intrusion} &    &  \cellcolor{LigthGray}TSA of SFTS   \cite{koumar2023} & \cellcolor[HTML]{BBFFBB}276 & \cellcolor[HTML]{BBFFBB}\textbf{98.48} & \cellcolor[HTML]{FAE5E4}\textbf{98.50} \\
               \multirow{-5}{*}{detection} &   \multirow{-3}{*}{UNSW-NB15 \cite{moustafa2015unsw}} & \cellcolor{LigthGray}NetTiSA   & \cellcolor[HTML]{BBFFBB}\textbf{52} &  \cellcolor[HTML]{BBFFBB}\textbf{99.22} & \cellcolor[HTML]{FAE5E4}88.79 \\
        
                \hline
                
                &   & \cellcolor{LigthGray}Dai et al.  \cite{dai2023glads} & \cellcolor[HTML]{BBFFBB}912 & \cellcolor{LigthGray}99.99 & \cellcolor[HTML]{FAE5E4}\textbf{99.65} \\
                \multirow{-2}{*}{TOR} &  &  \cellcolor{LigthGray}TSA of SFTS   \cite{koumar2023} & \cellcolor[HTML]{BBFFBB}276 & \cellcolor{LigthGray}99.84 & \cellcolor[HTML]{FAE5E4}96.33 \\
                \multirow{-2}{*}{detection} &  \multirow{-3}{*}{ISCX-Tor-2016 \cite{lashkari2017characterization}}  & \cellcolor{LigthGray}NetTiSA & \cellcolor[HTML]{BBFFBB}\textbf{52} & \cellcolor{LigthGray}99.91  &  \cellcolor[HTML]{FAE5E4}97.24 \\
                
                \hline
                
                 &   &  \cellcolor[HTML]{BBFFBB}Aceto  et  al.~\cite{ACETO2021102985} & \cellcolor[HTML]{BBFFBB}1,296 & \cellcolor[HTML]{BBFFBB}93.75  & \cellcolor[HTML]{BBFFBB}91.95   \\
                  &   &  \cellcolor[HTML]{BBFFBB}TSA of SFTS   \cite{koumar2023} & \cellcolor[HTML]{BBFFBB}276 & \cellcolor[HTML]{BBFFBB}94.35 & \cellcolor[HTML]{BBFFBB}\textbf{95.48} \\
                  &  \multirow{-3}{*}{ISCX-VPN-2016 \cite{icxs_vpn_2016_dataset}} &  \cellcolor[HTML]{BBFFBB}NetTiSA   & \cellcolor[HTML]{BBFFBB}\textbf{52} & \cellcolor[HTML]{BBFFBB}\textbf{98.64} & \cellcolor[HTML]{BBFFBB}92.64  \\     
        
                % \hline
                \cline{2-6}
                
                &   & \cellcolor[HTML]{BBFFBB}Jorgense et al.  \cite{vnat_dataset} & \cellcolor[HTML]{BBFFBB}3,612 & \cellcolor[HTML]{BBFFBB}-- & \cellcolor[HTML]{BBFFBB}98.00 \\
                \multirow{-5}{*}{VPN} &   &  \cellcolor[HTML]{BBFFBB}TSA of SFTS   \cite{koumar2023} & \cellcolor[HTML]{BBFFBB}276 & \cellcolor[HTML]{BBFFBB}\textbf{99.98} & \cellcolor[HTML]{BBFFBB}\textbf{99.73} \\
                \multirow{-5}{*}{detection} &  \multirow{-3}{*}{VNAT \cite{vnat_dataset}} & \cellcolor[HTML]{BBFFBB}NetTiSA   & \cellcolor[HTML]{BBFFBB}\textbf{52} & \cellcolor[HTML]{BBFFBB}\textbf{99.96} & \cellcolor[HTML]{BBFFBB}\textbf{99.35}  \\

                \hline
            \end{tabular}
            \label{tab:classification_results_binary}
        \end{center}
    \end{table}

\section{Confusion matrixes}
\label{sec:confusion_matrixes}

     \begin{table}[H]
            \caption{Confusion matrix for each binary classification problem. The first row represents the ``False'' labels, and the second row represents the ``True'' labels in the dataset. The first column represents predicted ``False'', and the second column is predicted ``True'' values by the classifier.}
            \begin{center}
                \begin{tabular}{RR|RR|RR|RR|RR}
                    \toprule
                    \multicolumn{2}{c}{\textbf{Botnet}} & \multicolumn{2}{c}{\textbf{Cryptomining}} & \multicolumn{2}{c}{\textbf{DoH -- RealWorld}} & \multicolumn{2}{c}{\textbf{DoH -- CIC}} & \multicolumn{2}{c}{\textbf{DNS Malware}} \\
                    \toprule
                    23,398 &  4 & 682,969 & 30 & 26,815 & 14,006 & 69,516 & 28 & 959 & 1 \\
                    2 & 1,596 & 28,812 & 363,765 & 2,103 & 457,076  & 78 & 30,378 & 5 & 35 \\
                    \bottomrule
                    \toprule
                    \multicolumn{2}{c}{\textbf{DoS}} & \multicolumn{2}{c}{\textbf{IoT -- Edge-IIoT}} & \multicolumn{2}{c}{\textbf{HTTPS Brute-force}} & \multicolumn{2}{c}{\textbf{IDS -- CIC}} & \multicolumn{2}{c}{\textbf{IDS -- UNSW}} \\
                    \toprule
                    984,942 & 0  & 232,541 & 4 & 95,663 & 3 & 287,752 & 546 & 561,766 & 2,099 \\
                    6 & 15,052  & 4 & 17,451 & 14 & 4,320 & 721 & 110,981 & 2,457 & 17,987 \\
                    \bottomrule
                    \toprule
                    \multicolumn{2}{c}{\textbf{TOR}} & \multicolumn{2}{c}{\textbf{IoT -- IoT-23}} & \multicolumn{2}{c}{\textbf{IoT -- TON\_IoT}} & \multicolumn{2}{c}{\textbf{VPN -- ISCX}} & \multicolumn{2}{c}{\textbf{VPN -- VNAT}} \\
                    \toprule
                    11,570 & 3  & 122,040 & 181 & 1,346 & 206 & 45,049 & 170 & 9,691 & 0 \\
                    7 & 176  & 550 & 377,229 & 52 & 498,396 & 509 & 4,272 & 4 & 305 \\
                    \bottomrule
                \end{tabular}
                \label{tab:confusion_binary}
            \end{center}
        \end{table}

        \begin{table}[H]
            \caption{Confusion matrix for multiclass classification of Botnet using CTU-13 dataset.}
            \begin{center}
                \begin{tabular}{cRRRRRRR}
                    \toprule
                      & & \multicolumn{6}{c}{\textbf{Predicted labels}} \\
                      & & \textbf{Clear} & \textbf{Donbot} & \textbf{Fast flux} & \textbf{Neris} & \textbf{Qvod} & \textbf{Rbot} \\
                    \toprule
                    & \textbf{Clear} & 23,428 & 0 & 0 & 0 & 0 & 0 \\
                    & \textbf{Donbot} & 0 & 3 & 2 & 0 & 0 & 0 \\
                    & \textbf{Fast flux} & 2 & 0 & 566 & 17 & 0 & 0 \\
                    & \textbf{Neris} & 1 & 0 & 11 & 923 & 0 & 0 \\
                    & \textbf{Qvod} & 0 & 0 & 1 & 0 & 38 & 0 \\
                    \multirow{-6}{*}{\rotatebox[origin=c]{90}{\textbf{True labels}}} & \textbf{Rbot} & 2 & 0 & 1 & 1 & 0 & 4 \\
                    \bottomrule
                \end{tabular}
                \label{tab:confusion_ctu_13}
            \end{center}
        \end{table}
        
        \begin{table}[H]
            \caption{Confusion matrix for multiclass classification of VPN using ISCX dataset.}
            \begin{center}
                \begin{tabular}{cRRRRRRRR}
                    \toprule
                    & & \multicolumn{6}{c}{\textbf{Predicted labels}} \\
                    & & \rotatebox[origin=c]{90}{\textbf{CHAT}} &
                    \rotatebox[origin=c]{90}{\textbf{EMAIL}} &
                    \rotatebox[origin=c]{90}{\textbf{FileTransfer}} &
                    \rotatebox[origin=c]{90}{\textbf{P2P}} &
                    \rotatebox[origin=c]{90}{\textbf{STREAMING}} &
                   \rotatebox[origin=c]{90}{\textbf{VOIP}} \\
                    \toprule
                    & \textbf{CHAT} & 904 & 0 & 33 & 0 & 3 & 48 \\
                    & \textbf{EMAIL} & 6 & 79 & 1 & 1 & 1 & 7 \\
                    & \textbf{FileTransfer} & 24 & 0 & 295 & 6 & 13 & 33 \\
                    & \textbf{P2P} & 3 & 0 & 3 & 139 & 0 & 6 \\
                    & \textbf{STREAMING} & 9 & 0 & 17 & 7 & 255 & 17 \\
                    \multirow{-6}{*}{\rotatebox[origin=c]{90}{\textbf{True labels}}} & \textbf{VOIP} & 29 & 0 & 24 & 2 & 5 & 3,030 \\
                    \bottomrule
                \end{tabular}
                \label{tab:confusion_vpn_iscx}
            \end{center}
        \end{table}

        \begin{table}[H]
            \caption{Confusion matrix for multiclass classification of VPN using VNAT dataset.}
            \begin{center}
                \begin{tabular}{cRRRRRR}
                    \toprule
                    & & \multicolumn{5}{c}{\textbf{Predicted labels}} \\
                    & & \textbf{C2} & \textbf{CHAT} & \textbf{FILE TRANSFER} & \textbf{STREAMING} & \textbf{VoIP} \\
                    \toprule
                    & \textbf{C2} & 3,906 & 2 & 3 & 16 & 0 \\
                    & \textbf{CHAT} & 9 & 918 & 3 & 6 & 0 \\
                    & \textbf{FILE TRANSFER} & 5 & 2 & 4,095 & 11 & 0 \\
                    & \textbf{STREAMING} & 8 & 2 & 2 & 862 & 0 \\
                    \multirow{-5}{*}{\rotatebox[origin=c]{90}{\textbf{True labels}}} & \textbf{VoIP} & 2 & 0 & 0 & 0 & 148 \\
                    \bottomrule
                \end{tabular}
                \label{tab:confusion_vpn_vnat}
            \end{center}
        \end{table}

        \begin{table}[H]
            \caption{Confusion matrix for multiclass classification of IoT Malware by using Edge-IIoTset dataset.}
            \begin{center}
                \begin{tabular}{cRRRRRRRRRRRRR}
                    \toprule
                     & & \multicolumn{12}{c}{\textbf{Predicted labels}} \\
                     & & \rotatebox[origin=c]{90}{\textbf{SQL injection}} & \rotatebox[origin=c]{90}{\textbf{XSS}} & \rotatebox[origin=c]{90}{\textbf{Backdoor}} & \rotatebox[origin=c]{90}{\textbf{Clear}} & \rotatebox[origin=c]{90}{\textbf{DDoS}} & \rotatebox[origin=c]{90}{\textbf{MitM}} & \rotatebox[origin=c]{90}{\textbf{OS fingerprinting}} & \rotatebox[origin=c]{90}{\textbf{Password attack}} & \rotatebox[origin=c]{90}{\textbf{Port scanning}} & \rotatebox[origin=c]{90}{\textbf{Ransomware}} & \rotatebox[origin=c]{90}{\textbf{Uploading attack}} & \rotatebox[origin=c]{90}{\textbf{Vulnerability scanner}} \\
                    \toprule
                    & \textbf{SQL injection} & 982 & 1 & 0 & 0 & 0 & 0 & 0 & 1 & 0 & 0 & 0 & 13 \\
                    & \textbf{XSS} & 0 & 354 & 0 & 0 & 1 & 0 & 0 & 0 & 0 & 0 & 0 & 0 \\
                    & \textbf{Backdoor} & 0 & 0 & 9 & 1 & 0 & 0 & 0 & 0 & 0 & 0 & 0 & 0 \\
                    & \textbf{Clear} & 0 & 0 & 0 & 229,064 & 2 & 0 & 0 & 0 & 0 & 0 & 0 & 0 \\
                    & \textbf{DDoS} & 0 & 0 & 0 & 2 & 2,358 & 0 & 0 & 1 & 0 & 0 & 0 & 2 \\
                    & \textbf{MitM} & 0 & 0 & 0 & 0 & 0 & 15 & 0 & 0 & 6 & 0 & 0 & 0 \\
                    & \textbf{OS fingerprinting} & 0 & 0 & 0 & 1 & 0 & 0 & 0 & 0 & 54 & 0 & 0 & 0 \\
                    & \textbf{Password attack} & 0 & 0 & 0 & 1 & 0 & 0 & 0 & 12,259 & 0 & 0 & 0 & 1 \\
                    & \textbf{Port scanning} & 0 & 0 & 0 & 1 & 1 & 0 & 0 & 0 & 3,551 & 0 & 0 & 0 \\
                    & \textbf{Ransomware} & 0 & 0 & 0 & 0 & 1 & 0 & 0 & 0 & 0 & 9 & 0 & 0 \\
                    & \textbf{Uploading attack} & 0 & 0 & 0 & 0 & 0 & 0 & 0 & 0 & 0 & 0 & 400 & 1 \\
                    \multirow{-12}{*}{\rotatebox[origin=c]{90}{\textbf{True labels}}} & \textbf{Vulnerability scanner} & 13 & 0 & 0 & 0 & 0 & 0 & 0 & 2 & 0 & 0 & 0 & 893 \\
                    \bottomrule
                \end{tabular}
                \label{tab:confusion_edge_iiot}
            \end{center}
        \end{table}

        \begin{table}[H]
            \caption{Confusion matrix for multiclass classification of IoT Malware using TON\_IoT dataset.}
            \begin{center}
                \begin{tabular}{cRRRRRRRRRR}
                    \toprule
                    & & \multicolumn{9}{c}{\textbf{Predicted labels}} \\
                    & & \rotatebox[origin=c]{90}{\textbf{Backdoor}} &
                    \rotatebox[origin=c]{90}{\textbf{Clear}} &
                    \rotatebox[origin=c]{90}{\textbf{DoS}} &
                    \rotatebox[origin=c]{90}{\textbf{Injection}} &
                    \rotatebox[origin=c]{90}{\textbf{MitM}} &
                    \rotatebox[origin=c]{90}{\textbf{Password}} &
                    \rotatebox[origin=c]{90}{\textbf{Runsomware}} &
                    \rotatebox[origin=c]{90}{\textbf{Scanning}} &
                    \rotatebox[origin=c]{90}{\textbf{XSS}} \\
                    \toprule
                    & \textbf{Backdoor}  & 3,928 & 10 & 153 & 55 & 13 & 199 & 151 & 240 & 216 \\
                    & \textbf{Clear}  & 7 & 719 & 14 & 9 & 1 & 6 & 3 & 31 & 10 \\
                    & \textbf{DoS}  & 230 & 9 & 44,732 & 173 & 11 & 358 & 56 & 266 & 311 \\
                    & \textbf{Injection}  & 63 & 3 & 268 & 27,794 & 4 & 400 & 25 & 132 & 629 \\
                    & \textbf{MitM}  & 47 & 2 & 54 & 6 & 76 & 26 & 24 & 55 & 49 \\
                    & \textbf{Password}  & 177 & 5 & 388 & 482 & 16 & 32,572 & 42 & 169 & 348 \\
                    & \textbf{Runsomware}  & 194 & 6 & 97 & 33 & 12 & 106 & 953 & 96 & 137 \\
                    & \textbf{Scanning}  & 450 & 10 & 317 & 121 & 19 & 204 & 51 & 21,867 & 778 \\
                    \multirow{-9}{*}{\rotatebox[origin=c]{90}{\textbf{True labels}}} & \textbf{XSS}  & 141 & 6 & 302 & 307 & 4 & 317 & 49 & 272 & 107,384 \\
                    \bottomrule
                \end{tabular}
                \label{tab:confusion_ton_iot}
            \end{center}
        \end{table}

        \begin{table}[H]
            \caption{Confusion matrix for multiclass classification of Intrusion in IDS system using CIC-IDS-2017 dataset.}
            \begin{center}
                \begin{tabular}{cRRRRRRRRRR}
                    \toprule
                     & & \multicolumn{9}{c}{\textbf{Predicted labels}} \\
                     & & \rotatebox[origin=c]{90}{\textbf{BENIGN}} & \rotatebox[origin=c]{90}{\textbf{Bot}} & \rotatebox[origin=c]{90}{\textbf{DDoS}} & \rotatebox[origin=c]{90}{\textbf{FTP-Patator}} & \rotatebox[origin=c]{90}{\textbf{Heartbleed}} & \rotatebox[origin=c]{90}{\textbf{Infiltration}} & \rotatebox[origin=c]{90}{\textbf{PortScan}} & \rotatebox[origin=c]{90}{\textbf{SSH-Patator}} & \rotatebox[origin=c]{90}{\textbf{Web Attack}} \\
                    \toprule
                    & \textbf{BENIGN} & 485,953 & 92 & 381 & 0 & 0 & 0 & 5 & 0 & 0 \\
                    & \textbf{Bot} & 213 & 500 & 0 & 0 & 0 & 0 & 0 & 0 & 0 \\
                    & \textbf{DDoS} & 251 & 0 & 134,320 & 0 & 0 & 0 & 2 & 0 & 3 \\
                    & \textbf{FTP-Patator} & 8 & 0 & 0 & 2,277 & 0 & 0 & 0 & 0 & 0 \\
                    & \textbf{Heartbleed} & 1 & 0 & 0 & 0 & 2 & 0 & 0 & 0 & 0 \\
                    & \textbf{Infiltration} & 13 & 0 & 0 & 0 & 0 & 0 & 0 & 0 & 0 \\
                    & \textbf{PortScan} & 20 & 0 & 219 & 0 & 0 & 0 & 224 & 1 & 0 \\
                    & \textbf{SSH-Patator} & 22 & 0 & 0 & 0 & 0 & 0 & 1 & 1,684 & 0 \\
                    \multirow{-9}{*}{\rotatebox[origin=c]{90}{\textbf{True labels}}} & \textbf{Web Attack} & 50 & 0 & 0 & 0 & 0 & 0 & 0 & 0 & 609 \\
                    \bottomrule
                \end{tabular}
                \label{tab:confusion_ids_cic}
            \end{center}
        \end{table}

        \begin{table}[H]
            \caption{Confusion matrix for multiclass classification of Intrusion in IDS system using UNSW dataset.}
            \begin{center}
                \begin{tabular}{cRRRRRRRRRRR}
                    \toprule
                    & & \multicolumn{10}{c}{\textbf{Predicted labels}} \\
                    & & \rotatebox[origin=c]{90}{\textbf{Analysis}} &
                    \rotatebox[origin=c]{90}{\textbf{Backdoor}} &
                    \rotatebox[origin=c]{90}{\textbf{Clear}} &
                    \rotatebox[origin=c]{90}{\textbf{DoS}} &
                    \rotatebox[origin=c]{90}{\textbf{Exploits}} &
                    \rotatebox[origin=c]{90}{\textbf{Fuzzers}} &
                    \rotatebox[origin=c]{90}{\textbf{Generic}} &
                    \rotatebox[origin=c]{90}{\textbf{Reconnaissance}} &
                    \rotatebox[origin=c]{90}{\textbf{Shellcode}} &
                    \rotatebox[origin=c]{90}{\textbf{Worms}} \\
                    \toprule
                    & \textbf{Analysis} & 0 & 0 & 89 & 1 & 11 & 0 & 0 & 0 & 0 & 0 \\
                    & \textbf{Backdoor} & 0 & 1,892 & 5 & 2 & 32 & 0 & 5 & 4 & 4 & 1 \\
                    & \textbf{Clear} & 0 & 14 & 563,592 & 13 & 341 & 1,352 & 41 & 141 & 31 & 0 \\
                    & \textbf{DoS} & 0 & 0 & 78 & 225 & 623 & 19 & 24 & 20 & 12 & 3 \\
                    & \textbf{Exploits} & 0 & 2 & 304 & 60 & 6234 & 35 & 105 & 156 & 36 & 13 \\
                    & \textbf{Fuzzers} & 0 & 0 & 2,358 & 4 & 41 & 3,228 & 3 & 3 & 1 & 0 \\
                    & \textbf{Generic} & 0 & 0 & 28 & 15 & 245 & 13 & 795 & 2 & 4 & 0 \\
                    & \textbf{Reconnaissance} & 0 & 84 & 19 & 2 & 281 & 5 & 2 & 2,907 & 0 & 0 \\
                    & \textbf{Shellcode} & 0 & 0 & 8 & 5 & 35 & 0 & 5 & 0 & 363 & 0 \\
                    \multirow{-10}{*}{\rotatebox[origin=c]{90}{\textbf{True labels}}} & \textbf{Worms} & 0 & 0 & 1 & 0 & 21 & 0 & 1 & 0 & 0 & 23 \\
                    \bottomrule
                \end{tabular}
                \label{tab:confusion_ids_unsw}
            \end{center}
        \end{table}

        \begin{table}[H]
            \caption{Confusion matrix for multiclass classification of TOR.}
            \begin{center}
                \begin{tabular}{cRRRRRRRRR}
                    \toprule
                    & & \multicolumn{8}{c}{\textbf{Predicted labels}} \\
                    & & \rotatebox[origin=c]{90}{\textbf{AUDIO}} &
                    \rotatebox[origin=c]{90}{\textbf{BROWSING}} &
                    \rotatebox[origin=c]{90}{\textbf{CHAT}} &
                    \rotatebox[origin=c]{90}{\textbf{FILE TRANSFER}} &
                    \rotatebox[origin=c]{90}{\textbf{MAIL}} &
                    \rotatebox[origin=c]{90}{\textbf{P2P}} &
                    \rotatebox[origin=c]{90}{\textbf{VIDEO}} &
                    \rotatebox[origin=c]{90}{\textbf{VOIP}} \\
                    \toprule
                    & \textbf{AUDIO} & 518 & 41 & 0 & 0 & 1 & 27 & 2 & 8 \\
                    & \textbf{BROWSING}  & 23 & 8,038 & 2 & 1 & 1 & 99 & 29 & 2 \\
                    & \textbf{CHAT}  & 4 & 26 & 63 & 1 & 0 & 6 & 7 & 7 \\
                    & \textbf{FILE TRANSFER}  & 2 & 7 & 0 & 566 & 0 & 13 & 3 & 0 \\
                    & \textbf{MAIL}  & 1 & 23 & 0 & 0 & 63 & 6 & 1 & 0 \\
                    & \textbf{P2P}  & 0 & 71 & 0 & 3 & 0 & 9,542 & 0 & 3 \\
                    & \textbf{VIDEO}  & 11 & 199 & 1 & 4 & 0 & 17 & 332 & 3 \\
                    \multirow{-8}{*}{\rotatebox[origin=c]{90}{\textbf{True labels}}} & \textbf{VOIP}  & 6 & 37 & 1 & 1 & 0 & 15 & 13 & 150 \\
                    \bottomrule
                \end{tabular}
                \label{tab:confusion_tor}
            \end{center}
        \end{table}

    % \newpage
\section{Description of merged multiclass dataset} \label{appendix:multiclass_multiclass}
    The complex multiclass dataset was created from following datasets: 1.\,CTU-13 \cite{GARCIA2014100}, 2.\,CESNET-MINER22 \cite{richard_plny_2022_7189293}, 3.\,CIC-Bell-DNS \cite{cic_bell_dns_2021_article}, 4.\,DoH-Real-world \cite{Jerabek2022}, 5.\,CIC-DoH-Brw \cite{montazerishatoori2020detection}, 6.\,Bot-IoT \cite{dos_iot_Dataset}, 7.\,HTTPS Brute-force  \cite{jan_luxemburk_2020_4275775}, 8.\,CIC-IDS-2017 \cite{sharafaldin2018toward}, 9.\,UNSW-NB-15 \cite{moustafa2015unsw}, 10.\,Edge-IIoTset \cite{9751703}, 11.\,IoT-23 \cite{sebastian_garcia_2020_4743746}, 12.\,TON\_IoT \cite{moustafa2021new}, 13.\,ISCX-Tor-2016 \cite{lashkari2017characterization}, and 14.\,ISCX-VPN-2016 \cite{icxs_vpn_2016_dataset}, and 15.\,VNAT \cite{vnat_dataset}. 

    In the created dataset, we merge classes with the same traffic resulting in the following 44 classes: Botnet-Neris, Clear, Botnet-RBot, Botnet-Fast\_flux, Botnet-Donbot, Botnet-Sogou, Botnet-Qvod, Cryptomining, DNS-malware, DoH, Backdoor, DoS, MitM, OS\_fingerprinting, Password\_attack, Scanning, Ransomware, Injection, Uploading\_attack, Vulnerability\_scanner, XSS, HTTPS-Brute-force, Bot, PortScan, Infiltration, FTP-Patator, SSH-Patator, Heartbleed, AUDIO, BROWSING, CHAT, File Transfer, EMAIL, P2P, VIDEO, VOIP, Fuzzers, Exploits, Shellcode, Worms, Reconnaissance, Backdoor, Analysis, and C2.

\section{Feature importance}
\label{sec:feature_importance}

        \begin{figure}[H]
            \centerline{\includegraphics[width=17cm]{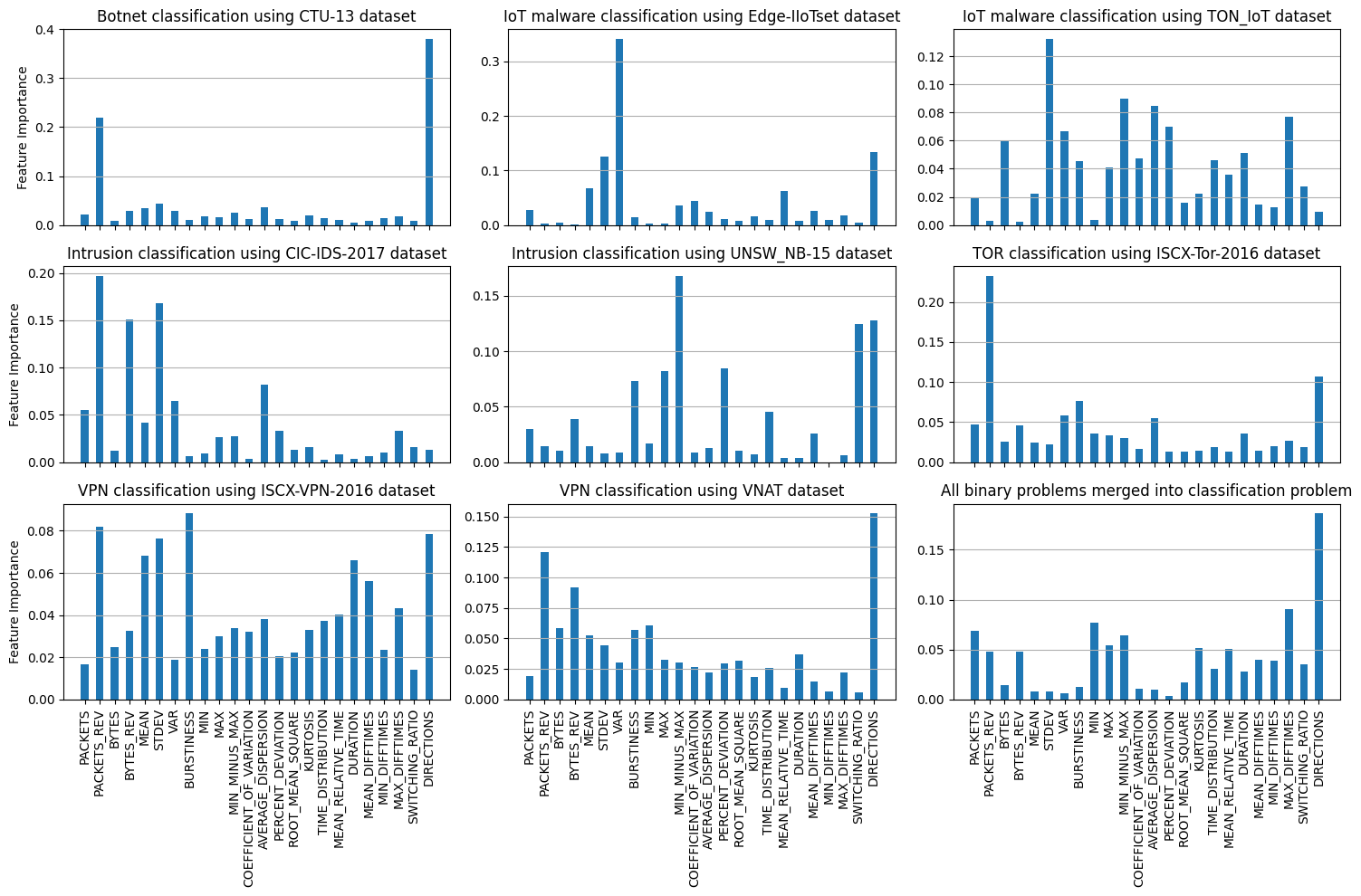}}
            \caption{Feature importance for multiclass classification}
            \label{fig:feature_imporance_multiclass}
        \end{figure}

         \begin{figure}[H]
            \centerline{\includegraphics[width=15cm]{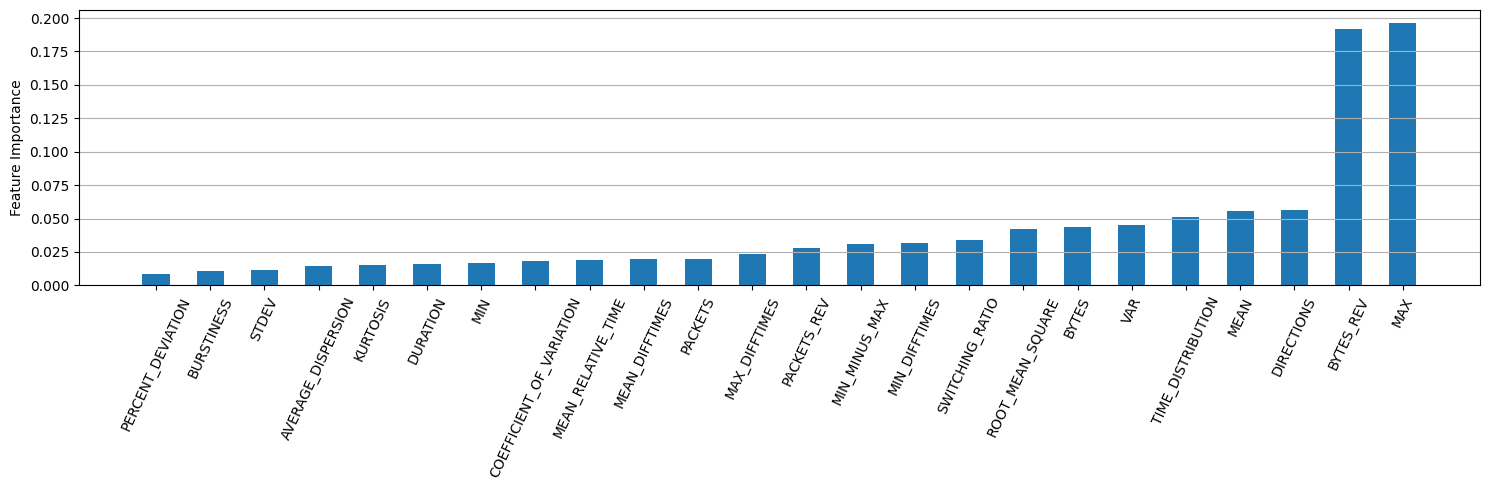}}
            \caption{Feature importance for classification on all dataset merged into multiclass problem with 44 classes.}
            \label{fig:feature_imporance_multiclass_all}
        \end{figure}

        \begin{figure}[H]
            \centerline{\includegraphics[width=17cm]{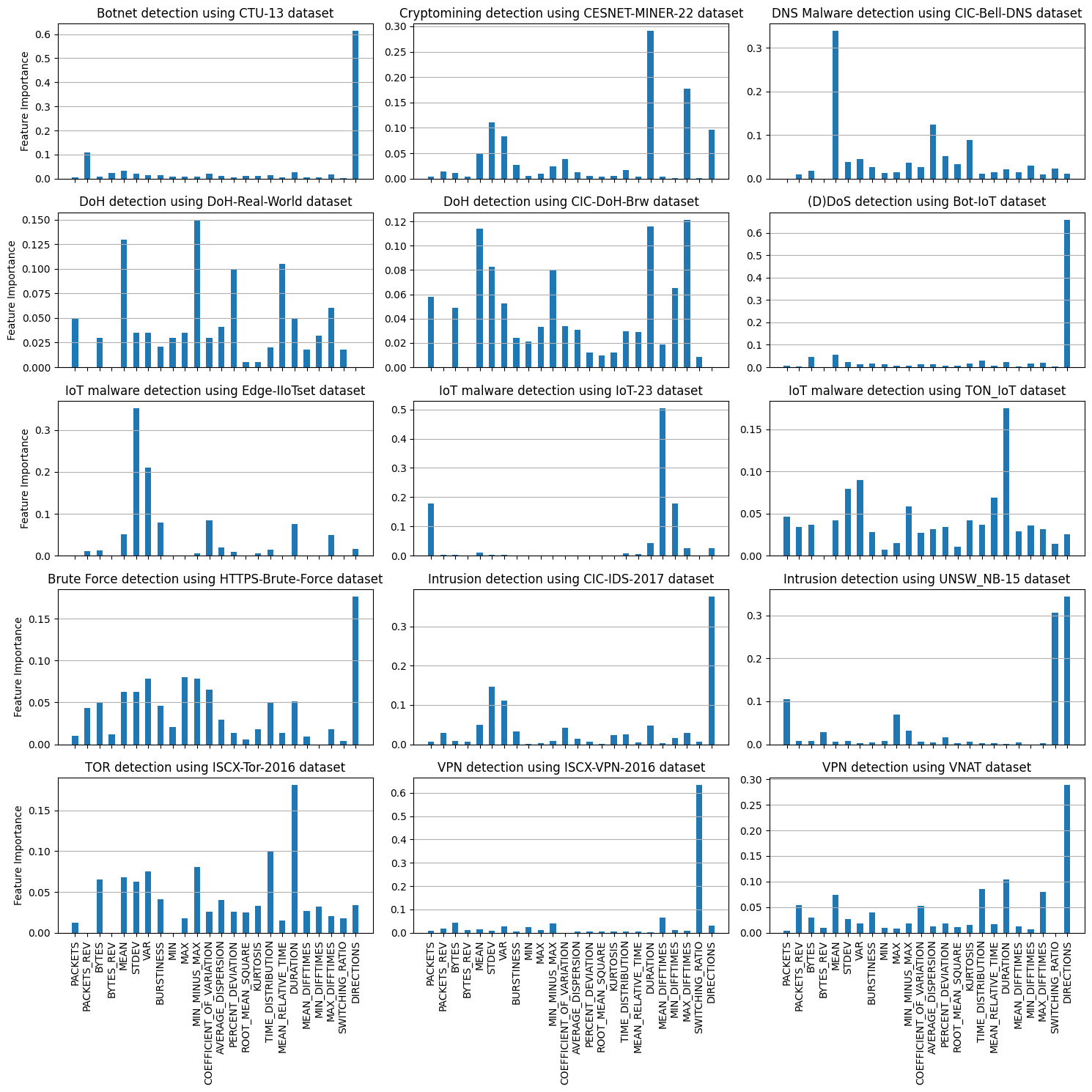}}
            \caption{Feature importance for binary classification}
            \label{fig:feature_imporance_binary}
        \end{figure}

    % \newpage

% \section{Correlation matrix} \label{correlation}
%     \begin{figure}[H]
%         \centerline{\includegraphics[width=15cm]{pictures/nettisa_correlation.png}}
%         \caption{Correlation matrix of Enhanced NetTiSA features.}
%         \label{fig:correlation_matrix}
%     \end{figure}

    % \newpage

%% References


\bibliographystyle{elsarticle-num}
\bibliography{bibliography.bib}

\end{document}